\documentclass[twocolumn]{aastex63}

\newcommand{\lya}{Ly$\alpha$}	
\newcommand{\zdla}{z_{\rm DLA}}	
\newcommand{\zqso}{z_{\rm QSO}}
\newcommand{\nhi}{N(H_I)}	
\newcommand{\lognhi}{\log N(H_I)}	
\newcommand{\meanflux}{\overline{f_\lambda}}	
\newcommand{\fluxunit}{10^{-19}\,{\rm W\,m^{-2}\,nm^{-1}}}	
\usepackage[squaren, Gray, cdot]{SIunits}

\received{June 1, 2019}
\revised{January 10, 2019}
\accepted{\today}
\submitjournal{ApJS}

\shorttitle{DLAs in SDSS DR16Q}
\shortauthors{Chabanier et al.}
\graphicspath{{./}{figures/}}

\begin{document}

\title{The Completed SDSS-IV extended Baryon Oscillation Spectroscopic Survey: The Damped Lyman-$\alpha$ systems Catalog}

\correspondingauthor{Sol\`ene Chabanier}
\email{schabanier@lbl.gov}

\author[0000-0002-5692-5243]{Sol\`ene~Chabanier}
\affiliation{Lawrence Berkeley National Laboratory, Berkeley, CA 94720}
\affiliation{IRFU, CEA, Universit\'e Paris-Saclay, F91191 Gif-sur-Yvette, France}

\author{Thomas~Etourneau}
\affiliation{IRFU, CEA, Universit\'e Paris-Saclay, F91191 Gif-sur-Yvette, France}

\author{Jean-Marc~Le Goff}
\affiliation{IRFU, CEA, Universit\'e Paris-Saclay, F91191 Gif-sur-Yvette, France}

\author{James~Rich}
\affiliation{IRFU, CEA, Universit\'e Paris-Saclay, F91191 Gif-sur-Yvette, France}

\author{Julianna~Stermer}
\affiliation{Sorbonne Universit\'e, Universit\'e Paris Diderot, CNRS/IN2P3, Laboratoire de Physique Nucl\'eaire et de Hautes Energies, LPNHE, 4 Place Jussieu, F-75252 Paris, France}

\author{Bela~Abolfathi}
\affiliation{Department of Physics and Astronomy, University of California, Irvine, CA 92697, USA}

\author{Andreu~Font-Ribera}
\affiliation{Institut de Física d’Altes Energies, The Barcelona Institute of Science and Technology, Campus UAB, 08193 Bellaterra (Barcelona), Spain}

\author{Alma X. Gonzalez-Morales}
\affiliation{Consejo Nacional de Ciencia y Tecnolog\'ia, Av. Insurgentes Sur 1582. Colonia Credito Constructor, Del. Benito Jurez C.P. 03940, M\'exico D.F. M\'exico}
\affiliation{Departamento de F\'isica, Divisi\'on de Ciencias e Ingenier\'ias, Campus Leon, Universidad de Guanajuato, L\'eon 37150, M\'exico }

\author{Axel~de~la~Macorra}
\affiliation{Universidad Nacional Aut\'onoma de M\'exico Instituto de Física Apdo. Postal 20-364, M\'exico}

\author{Ignasi P\'erez-R\'afols}
\affiliation{Sorbonne Universit\'e, Universit\'e Paris Diderot, CNRS/IN2P3, Laboratoire de Physique Nucl\'eaire et de Hautes Energies, LPNHE, 4 Place Jussieu, F-75252 Paris, France}

\author{Patrick~Petitjean}
\affiliation{Institut d'Astrophysique de Paris, Sorbonne Universit\'es and CNRS,
   98bis Boulevard Arago, 75014, Paris, France}	
 
   \author{Matthew~M.~Pieri}
\affiliation{Aix Marseille Universit\'es, CNRS, CNES, Laboratoire d'Astrophysique de Marseille, Marseille, France}
   
 \author{Corentin~Ravoux}
\affiliation{IRFU, CEA, Universit\'e Paris-Saclay, F91191 Gif-sur-Yvette, France}	

 \author{Graziano~Rossi}
\affiliation{Department of Physics and Astronomy, Sejong University, Seoul, 143-747, Korea}	

 \author{Donald P. Schneider}
\affiliation{Department of Astronomy and Astrophysics, The Pennsylvania State University, University Park, PA 16802, USA}	
\affiliation{Institute for Gravitation and the Cosmos, The Pennsylvania State University, University Park, PA 16802, USA}



\begin{abstract}

  We present the characteristics of the Damped Lyman-$\alpha$ (DLA) systems
  found in the data release DR16 of the
  extended Baryon Oscillation Spectroscopic Survey (eBOSS) of the
  Sloan Digital Sky Survey (SDSS).
  DLAs were identified using the convolutional neural network (CNN)
  of~\cite{Parks2018}.
  A total of 117,458  absorber candidates were found with $2 \leq \zdla \leq 5.5$ and $19.7 \leq \lognhi \leq  22$, including 57,136 DLA candidates with $\lognhi \geq  20.3$.
  Mock quasar spectra were used to estimate DLA detection efficiency
  and the purity of the resulting catalog.
  Restricting the quasar sample to bright forests, i.e. those with
  mean forest fluxes $\meanflux>2\times\fluxunit$,
  the completeness and purity are greater than 90\% for
  DLAs with column densities
  in the range $20.1\leq \lognhi \leq 22$. 
\end{abstract}



\section{Introduction} \label{sec:intro}

Damped Lyman-$\alpha$ (DLA) are absorption systems with neutral hydrogen column densities, $\nhi$ $\geq 2 \times 10^{20}$ atoms/cm$^{2}$, producing broad damping wings in the optical spectra of bright background objects such as quasars \citep{Wolfe1986}.

Such systems are at high enough density to be self-shielded against ionizing radiation \citep{Vladilo2001, Cen2012,Fumagalli2014} and they are connected to dark matter halos over a large range of masses, from dwarf galaxies to cluster of galaxies \citep{Haehnelt1998,Prochaska1997,Pontzen2008}.
Observations show that DLAs are the dominant reservoir of neutral hydrogen in the redshift range $0<z< 5$ and contain 2\% of all baryons in the universe \citep{Gardner1997, Wolfe2005, Prochaska2009,Noterdaeme2012}. As such, DLAs are keys to understanding galaxy formation and evolution  since they are thought to be the reservoir of atomic gas for stellar formation in galaxies.
They are thus an important probe of physical conditions in the interstellar medium at high redshifts \citep{Petitjean2000,Fumagalli2013,Bird2014,Ota2014,Fumagalli2016,Rudie2017}.

But DLAs are also contaminants in the measurements of the Lyman-$\alpha$ (\lya) forest  flux probability distribution function \citep{Lee2015}, its 3D auto-correlation function \citep{Slosar2011, Bautista2017, dMdB2020} or its 1D power spectrum \citep{McDonald2006,PalanqueDelabrouille2013,Chabanier2019}. Since DLAs form at high density peaks they cluster more strongly than diffuse \lya\ clouds \citep{FontRibera2012}, thus biasing astrophysical and cosmological parameters if not well accounted for. Therefore, their detection along with the  measurements of their physical properties, absorption redshift and column densities, are important in such studies.

With hundreds of thousands detected quasar spectra, the large statistical power of the Sloan Digital Sky Survey (SDSS, \cite{York2000}) has fostered the compilation of DLA catalogs \citep{Noterdaeme2009,Prochaska2009, Zhu2013,Garnett2017,Parks2018,Ho2020}.
Given the tremendous number of spectra to analyze, it has also played a critical role for the development of automated detection algorithms over visual inspection, e.g. using  Voigt-profile fitting \citep{Prochaska2005,Noterdaeme2009,Noterdaeme2012} or machine learning techniques such as convolutional neural networks (CNN,\cite{Parks2018}), Gaussian processes \citep{Garnett2017} or random forest classifiers \citep{Fumagalli2020}.

The final SDSS-IV quasar catalog from Data Release 16 (DR16) of the extended Baron Oscillation Spectroscopic Survey (eBOSS, \cite{Dawson2016,SDSSDR16}), which we will refer to as DR16Q, is the largest quasar spectra sample to date with 920,110 observations of 750,414 quasars \citep{Lyke2020}. In the DR16Q, we used the CNN algorithm from \cite{Parks2018} to include DLA quasar identification for very confident DLAs with $\lognhi$ $\geq 20.3$ only.
Here we present the full sample of absorbing systems detected with the CNN in DR16Q, which includes less confident DLAs and Lyman Limit Systems (LLS) with $\log(\nhi)$ as low as 19.7. The choice of the CNN from \cite{Parks2018} is motivated by the design of the algorithm constructed specifically for low redshift and low signal to noise BOSS/eBOSS quasar spectra.

The paper is organized as follow. Sec.~\ref{sec:dr16q} presents the quasar spectra sample which is scanned for high column density absorbing systems. Sec.~\ref{sec:cnn} introduces the automated algorithm and the CNN architecture from \cite{Parks2018} that we use to detect strong absorbers. We perform efficiency and purity validation of the algorithm with synthetic spectra and a study of  biases  of DLA parameters, $\log(\nhi)$ and $\zdla$ in Sec.~\ref{sec:tests}. Finally, we present the full absorber sample in Sec.~\ref{sec:results} and compare it with existing catalogs. We present concluding remarks in Sec.~\ref{sec:conclusion}.

\section{Quasar spectra sample DR16Q}
\label{sec:dr16q}
In this work, we use data measured with BOSS and eBOSS \citep{Dawson2016}  of the SDSS-III and SDSS-IV \citep{Gunn2006,Smee2013,Blanton2017} surveys respectively.
We focus on the \lya\ forest regions from the 750,414 quasar spectra available in DR16Q \citep{Lyke2020}, which contains all SDSS spectroscopically observed quasars. The selection of quasars for the BOSS and eBOSS surveys are described in~\cite{Ross2012,Myers2015}.

We search for DLAs in the 263,201
spectra with $2 \leq \rm{{\sc Z\_LYAWG}} \leq 6 $, the redshift range over which spectra contain enough pixels to identify DLAs.
We use the quasar redshift estimator, {\sc Z\_LYAWG}, generated by Principal Component Analysis (PCA), using the {\sc redvsblue} algorithm\footnote{\url{https://github.com/londumas/redvsblue}}.
The DR16Q catalog is constructed from the {\sc spAll-v5\_13\_0} (spAll) file containing all SDSS-III/IV observations treated by the version {\sc v5\_13\_0} of the SDSS spectroscopic pipeline\footnote{\url{https://data.sdss.org/datamodel/files/BOSS_SPECTRO_REDUX/RUN2D/spAll.html}}. If multiple observations are available for one object in the spAll file,  we use the stacked spectrum
of all good observations as input to the DLA finder.
We identify bad spectra using the ZWARNING parameter. If ZWARNING was SKY, LITTLE COVERAGE, UNPLUGGED, BAD TARGET, or NODATA, we did not use the associated observation in the stack.

   \begin{figure}
   \centering
   \includegraphics[width=\columnwidth]{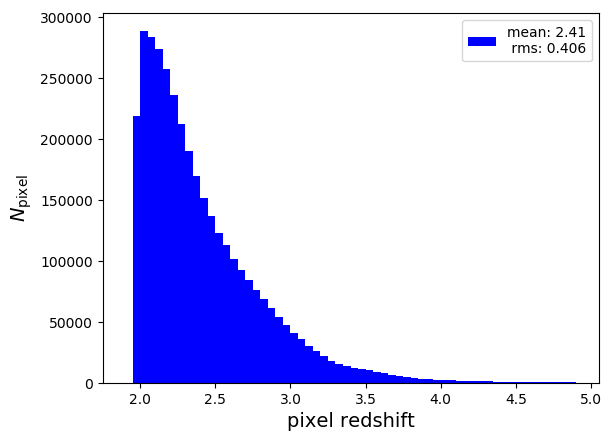}
   \caption{Redshift distribution for
     pixels in the \lya~forest  from quasar spectra available in DR16Q.
}
         \label{fig:zpixel}
   \end{figure}
   
   \begin{figure}
   \centering
   \includegraphics[width=\columnwidth]{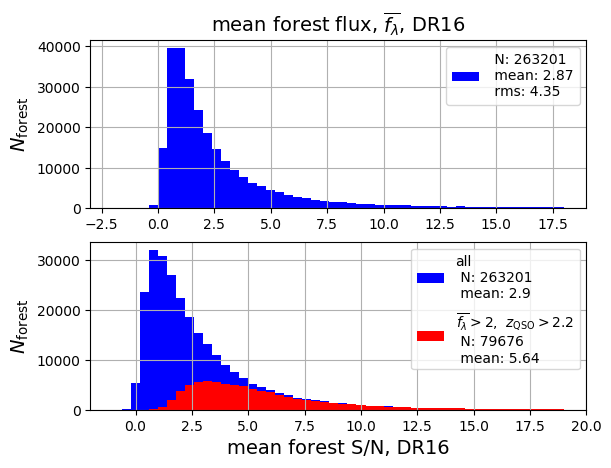}
   \includegraphics[width=\columnwidth]{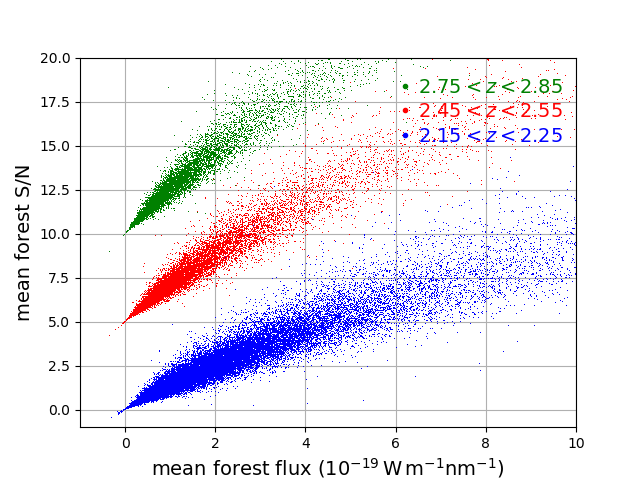}
   \caption{The mean flux and mean
     signal-to noise ratio (S/N) for pixels in the \lya~forest.
     \textbf{Top panel:} shows the distribution for the complete sample (blue)
     and for the restricted sample with high DLA detection efficiency:
     quasar redshift $\zqso>2.2$ and mean forest flux $\meanflux>2.0$.
     \textbf{Bottom panel:} shows S/N as a function of  $\meanflux$
     for three ranges of mean forest redshift as labeled. For
     clarity the S/N is offset by (0,5,10) units.
}
         \label{fig:snr}
   \end{figure}

   Fig.~\ref{fig:zpixel} shows the redshift distribution of the forest pixels,
   with a mean of $z=2.4$.
   Fig.~\ref{fig:snr} shows the flux and signal-to-noise (S/N) averaged over
   the forest, with a mean of $2.87\times\fluxunit$ and 2.90, respectively.
   We will see in Sect.~\ref{sec:effpure} that the efficiency for
   finding DLAs is poor for forests with low S/N corresponding
   generally to forests with low fluxes.
   Fig.~\ref{fig:snr} therefore also shows the S/N for a ``bright
   sample'' of forests with
   mean forest flux $\meanflux>2\times\fluxunit$.
   Also shown is the S/N as a function of $\meanflux$ for three
   redshift ranges.
   We see that bright forests generally have S/N greater than 2.

\section{DLA detection method}
\label{sec:cnn}

We identified DLAs with the algorithm described in~\cite{Parks2018}, which is based on a multi-task learning CNN. We refer the reader to~\cite{Parks2018} for a complete description of the detection algorithm, only  recalling here the major steps. The CNN architecture  and its training aim at constructing an algorithm that works at low redshifts, in noisy regions, and without any  input from the user other than raw spectral data.  The algorithm therefore does not need quasar continuum or DLA Voigt profile modeling and it ignores flux errors estimated by the SDSS pipeline. Finally, the model does not include Broad Absorption Lines (BALs), compromising DLA detection. Therefore we reject lines of sight that the DR16Q pipeline indicates as affected by BALs.

The neural network model uses a standard 2D CNN architecture with four layers. It relies on the Adam (Adaptive Moment Estimation) algorithm to search for the optimal parameters \citep{Adam2014} and is implemented using the Google's deep learning framework TensorFlow\footnote{\url{https://www.tensorflow.org/}}. It analyzes 1,748-pixel long sightlines  of  $\delta \lambda \simeq 1 \ \angstrom$  in 1,748 inference steps with 400-pixel long sliding windows in the $900 \ \angstrom \leq \lambda \leq 1,346 \ \angstrom$ region  in order to improve detection of multiple DLAs per-sightline. The 400-pixel size is in part imposed by the SDSS resolution. The model produces three outputs for each sliding window: (1) classification of the segment as containing a DLA or not, (2) the DLA absorption redshift {\sc Z\_DLA}, i.e. the pixel center localization and (3) the HI column density, {\sc NHI\_DLA}, if a DLA is visible. In the case of a detected DLA in a sightline, the authors also define a non statistical measure of confidence, the \textit{confidence} parameter over the range (0,1). It is based on how robustly the DLA is localized over the different predictions of the sliding window.

The training sample was constructed using 4,113 SDSS sightlines, with quasar redshift $z_{\rm qso} > 2.3$ and signal to noise  S/N$> 5$, identified as DLA-free from the analysis of~\cite{Prochaska2009}. The authors generated 200,000 sightlines from the DLA-free sample by inserting DLAs and super Lyman-limit systems (SLLS)
with logarithmic column density $19.5 \leq \lognhi \leq 22.5$ using Voigt profile modeling.

Finally, the algorithm was validated using one catalog with synthetic DLA in real DLA-free spectra and one catalog constituted of visually inspected spectra containing DLAs~\citep{Prochaska2009}.
The authors found a systematic bias
of order $\sim 0.1$ in the predicted $\lognhi$ at both low and high ends.
They  fit this bias with a 3$^{rd}$ degree polynomial
(see Fig.9 of~\cite{Parks2018})
and used this result to correct for the bias in the final automated algorithm.

\section{Analysis of DLAs in mock spectra}
\label{sec:tests}

Given that S/N and quasar redshift distributions of the training and validation samples do not exactly match those of DR16 data,
we used synthetic spectra to perform purity and efficiency validation of the algorithm along with an investigation of systematics on the inferred $\zdla$ and $\nhi$.
The synthetic spectra, hereafter ``mocks'', were  produced for the eBOSS Ly$\alpha$ data analysis~\citep{dMdB2020}. In Sec.~\ref{sec:mocks} we briefly describe the construction of mock spectra and we present our
estimates of efficiency and  purity in Sec.~\ref{sec:effpure}.
The accuracy and precision of the estimation of $\nhi$ is discussed
in Sec.~\ref{sec:recovery}.

\subsection{Synthetic quasar spectra}
\label{sec:mocks}

The production of the mocks is described elsewhere~\citep{Etourneau2021}, and
we describe here only the major steps.
A low resolution Gaussian random density field was produced in a box of $2560\times2560\times1536$ voxels of 2.19 $h^{-1}$ Mpc sides.
Quasar positions were drawn proportionally to a lognormal field, with phases in Fourier space equal to those of the density field.
The interpolated values of the density field along the sightlines from the observer position towards the quasars were computed, together with the 3 components of the velocity and the 6 components of the velocity gradient tensor.  Extra small-scale fluctuations were added to each sightline independently, in order to reproduce the variance in the \lya\ forest in the data. 
Redshift space distortions were implemented by adding the velocity gradient along the sightlines to the density field.
We then applied a log-normal transformation to this sum, and used the Fluctuating Gunn-Peterson Approximation (FGPA) to compute the optical depth in each cell. 

We define high column density systems (HCD) as systems with column density, $\nhi > 10^{17.2}$,  including both DLA and Lyman limit systems $10^{17.2} < \nhi <  2 \times 10^{20}$.
We selected peaks of the large-scale density field as possible locations of HCD and set the threshold to get a constant bias $b_{\rm HCD}(z)=2$ for the HCD, using formulas in appendix A of~\citep{FontRibera2012}. 
We then Poisson sampled the selected peaks to follow the HCD redshift 
distribution of the default model from the IGM physics package pyigm\footnote{https://github.com/pyigm/pyigm}.
The column density was selected to follow the same model. These distributions are illustrated in Fig.~\ref{fig:distrib_dla}.
The radial velocity of the HCD was obtained from the three velocity-component boxes. This information was included in a HCD catalog.

In a last step, the quasar spectra are produced by multiplying the transmitted flux fraction by a quasar continuum and adding instrumental noise (Gonzales-Morales et al. (in preparation)). For each HCD in the catalog, we multiplied the corresponding quasar spectrum by the Voigt profile for the HCD column density.

Figure \ref{fig:mnflux_mocks} shows the mean flux and mean S/N for mock
pixels in the \lya~forest.
We see that the mock distribution agree qualitatively with the
data, shown in Fig.~\ref{fig:snr}.
The data do, however,
contain more forests with very low flux and very high flux.

\subsection{Efficiency and purity}
\label{sec:effpure}

\begin{figure}[t]
   \centering
       \includegraphics[width = 0.47\columnwidth]{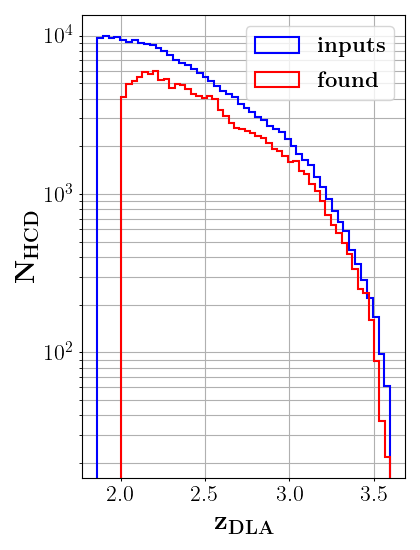}
    \includegraphics[width = 0.47\columnwidth]{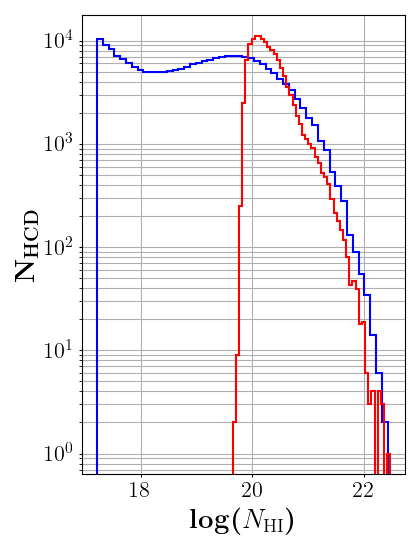}
    \caption{Mock HCD distribution of $\zdla$ (left) and $\log \nhi$ (right) as placed in mocks (blue) and recovered by the CNN (red). Mock spectra have a total of 218,124 HCDs. Among them, 91,659 HCDs have  $\log \nhi \geq 19$ and $\zdla \geq 2$. The CNN detects 132,226 absorbing systems with $\log \nhi \geq 19.6$ and $\zqso>\zdla \geq 2$. The CNN has not been trained for absorbers with $\log \nhi < 19.5$ but it still detects HCDs with $\log \nhi$ slightly below this threshold value, which explains the excess at the low end of the $\log \nhi$ distribution for the CNN.
    }
         \label{fig:distrib_dla}
\end{figure}

\begin{figure}
    \includegraphics[width = \columnwidth]{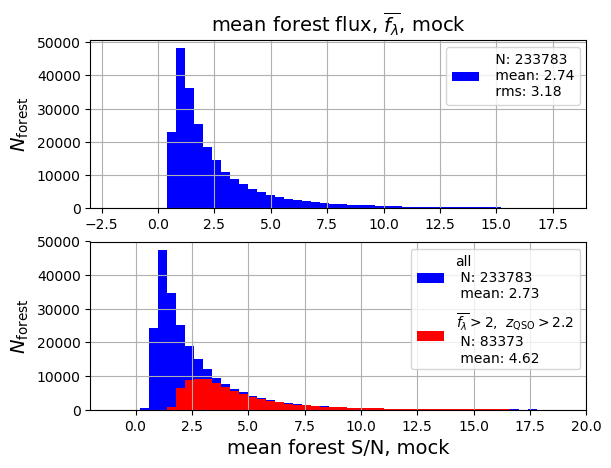}
    \caption{Mean flux and mean S/N for mocks pixels in the \lya-forest for the complete sample (blue) and for the restricted sample with high DLA detection efficiency (red): quasar redshift $\zqso>2.2$ and mean forest flux $\meanflux>2.0$.}
    \label{fig:mnflux_mocks}
    \end{figure}

The efficiency for DLA-detection and the purity of the detected sample
were studied by using the mock spectra where the catalog of
detected DLAs can be compared with the catalog of generated
HCDs. The model reports a total of 132,226
candidate absorbing systems with $\zdla < \zqso$ and $\zdla \geq 2$ in 92,042 sightlines with $\log \nhi$ and $\zdla$ distributions shown in Fig~\ref{fig:distrib_dla}.

Both efficiency and purity are functions of the characteristics of the forest (signal-to-noise ratio and forest mean flux) and of the DLA (redshift and column density).
They also depend on the criterion used to define
detected DLAs,
i.e.  the requirements placed on the confidence parameter
and on the required agreement between  generated and found $\zdla$ and $\nhi$.

The criterion for detection and matching are arbitrary to a certain extent.
The most important matching criterion concerns the redshift difference
between generated and detected DLAs.
Fig.~\ref{fig:dz_vs_continuum} shows this difference
vs the mean forest flux, $\meanflux$ for best matched DLAs,
where the match only requires that mock and found DLAs are on
the same sightline.  
Here, we adopt a matching criterion
requiring that detected and generated redshifts differ
at most by $\Delta z<0.02$ (about $25 \angstrom$).
This redshift-matching
cut accepts most detected DLAs for 
$\meanflux>2\times\fluxunit$. However, the redshift resolution
degrades substantially for lower $\meanflux$.
With the above criterion, the DLA finder recovers 62,847 absorbing systems with $\zdla > 2$ and $\log \nhi > 19$ (69\% of the absorbing systems put in mocks). Among them 86\% (70\%) have confidence parameters $>0.5$ ($>0.9$). 
Changing the redshift-matching criterion to $\Delta z<0.01$ reduces only slightly the number of recovered DLAs  from 63,847 to 61,131.

For the adopted matching criterion, $\Delta\zdla<0.02$,
the efficiency and purity as a function of  $\zdla$ and $\nhi$ are shown in
Fig.~\ref{fig:eff_pur_vs_n_z} on the left and right panels respectively.  Fig.~\ref{fig:eff_pur_vs_n_z_bright} and Fig.~\ref{fig:eff_pur_vs_n_z_faint} show the same measurements but for bright ($\meanflux > 2 \times\fluxunit$) and faint ($\meanflux > 2 \times\fluxunit$) forests respectively. Note that we use HCD characteristics as returned by the finder to compute the purity and the ones from the mock input for the efficiency, which explains why the right panel does not have data for $\log \nhi < 19.65$ but the left one does.

For the bright sample, Fig.~\ref{fig:eff_pur_vs_n_z_bright} shows that
high efficiency ($>0.9$) \emph{and} purity ($>0.9$) is obtained
for column densities in the range $20.2<\lognhi<22.0$ and redshifts $\zdla>2.2$.
For the faint sample, high efficiency and purity is found only for
$\lognhi>21.0$ and $\zdla>2.2$.

 For the efficiency, there is almost no dependence with $\zdla$. It is degraded for $\zdla \leq 2.2$ but performs quite equally for higher redshifts.
The bad performances at low absorbing redshifts happen since they occur in the blue and noisy end of the spectra (see S/N distribution as a function of quasar redshift in the bottom panel of Fig~\ref{fig:snr}). Indeed, false negatives have a mean forest flux 25\%  lower than the average. By comparing Figs.~\ref{fig:eff_pur_vs_n_z}, \ref{fig:eff_pur_vs_n_z_bright}, and \ref{fig:eff_pur_vs_n_z_faint} we easily deduce that faint forests are driving the bad performances. Also, because the spectra is small in size at low redshifts, i.e. with a low number of pixels, it is harder for the CNN to detect features and to do accurate predictions.
The efficiency drops below 0.2 for the low-end of $\nhi$, for which the CNN has not been specifically designed and trained 
  and for which  instrumental noise and resolution  make detection difficult.
  The finder detects HCDs with $\log \nhi$ as low as 19 but, as we 
will see in the next section, 
  overestimates this parameter. This explains the excess of the detected $\lognhi$ near the detection threshold compared to the mock distribution on the left panel of Fig.~\ref{fig:distrib_dla} 
 
  The efficiency also decreases for high $\nhi$ where the DLA covers a substantial fraction of the forest. While we observe a trend for a decrease toward the high-end of $\nhi$, synthetic spectra have a total of 806 HCD with $\log \nhi > 21.5$.
  While this makes our results statistically significant for high-$\nhi$ on average, results can be very noisy when sampled into $\zdla$ bins.
  Over the 104 missed DLAs with $\log \nhi > 21.5$ and $\zdla > 2.2$, 19 are detected by the finder but rejected by the redshift-matching cut criterion ($0.02<\Delta z<0.04$) because of low mean forest flux ($\meanflux<2$).
  Four have an absorbing redshift extremely close to the Ly$\alpha$ emission line such that the CNN found a $\zdla >\zqso$, 15 are part of two overlapping DLAs with $\Delta \zdla < 0.03$ detected as one DLA with a higher $\nhi$ (as was noted in~\cite{Parks2018}, the CNN struggles at identifying overlapping DLAs).
  The 66 remaining have particularly low mean forest fluxes with an average of $\sim 0.3 \times\fluxunit$. When considering bright forests only, with $\meanflux > 2\times\fluxunit$, the efficiency is alway $>0.9$ for $20<\lognhi$. The results are noisy for $\lognhi > 21.5$, especially for high-redshift bins, but the efficiency is close to one on average for $\lognhi > 21.5$ DLAs.
 
The purity is $>0.5$ for $\zdla<3.2$ and  $20.3<\lognhi < 21.5$, and  $>0.9$ for $\zdla<3.2$ and  $20.8<\lognhi < 21.5$. 

Our matching criterion does not use the confidence parameter and using it could increase the purity at the cost of decreasing efficiency.
Fig.~\ref{fig:conf_hist} shows the distribution of the confidence parameter for all found HCDs, true positives and false positives.
{False positives are all HCDs found by the CNN but that has not been matched to a HCD input (same sightline with $\Delta \zdla < 0.02.$)
Only 18\% (44\%) of false positives are \textit{confident} HCDs with confidence parameter $> 0.9$ ($>0.5$). Taking only HCDs with confidence parameter $>0.9$ brings the purity to be always $>0.9$ for $\log \nhi > 20.3$.

The net decrease of the purity toward low $\log \nhi$ seen in the right panel of Fig.~\ref{fig:eff_pur_vs_n_z} occurs since it gets more and more difficult for the CNN to distinguish between real but relatively small absorptions and noise. Indeed, when considering bright forests only (see right panel of Fig~\ref{fig:eff_pur_vs_n_z_bright}), with $\meanflux > 2\times\fluxunit$, the purity is alway $>0.9$ for $20.1<\lognhi$. We observe a decrease of purity at high redshifts
for both bright and faint samples (see Figs.~\ref{fig:eff_pur_vs_n_z},\ref{fig:eff_pur_vs_n_z_bright},\ref{fig:eff_pur_vs_n_z_faint}) since the mean flux decreases for high redshift quasar making it harder for the CNN to distinguish between Ly$\alpha$ absorptions and DLAs.

To summarize, the main parameter for maximal completeness and purity of the absorber catalog is the mean flux of the forest. Taking $\meanflux>2$ ensures efficiency and purity to be $>0.9$ for $\lognhi > 20.1$. However, it degrades the size of the sample. If taking all bright and faint forests, the efficiency is $>0.9$ for $\zdla > 2.2$ and  $20<\lognhi<21.5$, and purity is $> 0.9$ for $\zdla<3.2$ and  $20.5<\lognhi < 21.5$ and confidence $>0.9$.

\begin{figure}[t]
   \centering
    \includegraphics[width = \columnwidth]{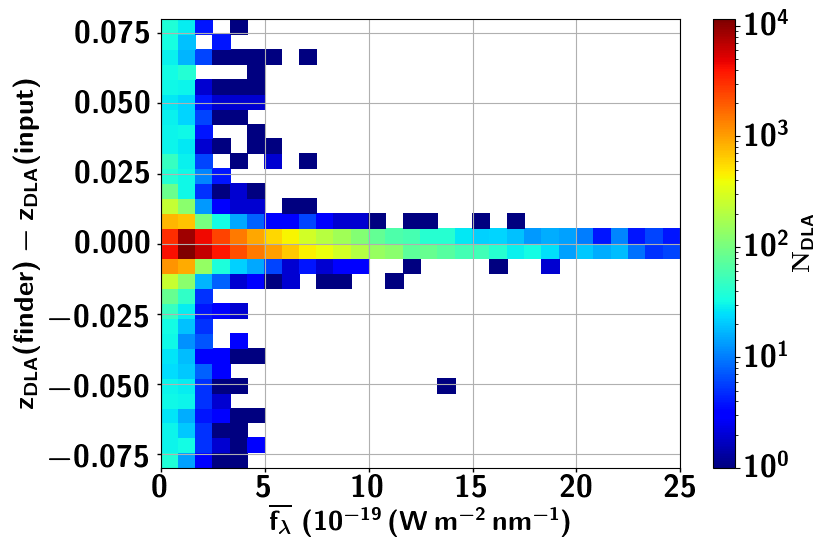}
   \caption{Mean difference between detected and generated DLA redshifts
     vs the mean forest flux, $\meanflux$.
   }
         \label{fig:dz_vs_continuum}
   \end{figure}

\begin{figure}
   \centering
   \includegraphics[width = \columnwidth]{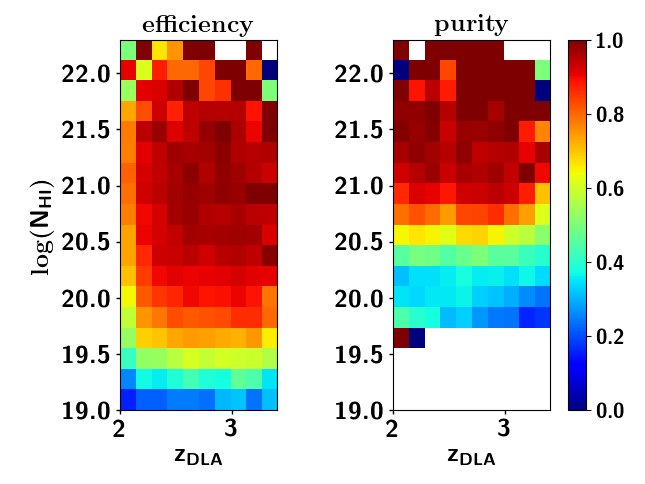}
   \caption{Efficiency (left) and purity (right) vs. redshift and column density using matching criterion $\Delta \zdla < 0.02$.
   }
         \label{fig:eff_pur_vs_n_z}
\end{figure}

\begin{figure}
   \centering
   \includegraphics[width = \columnwidth]{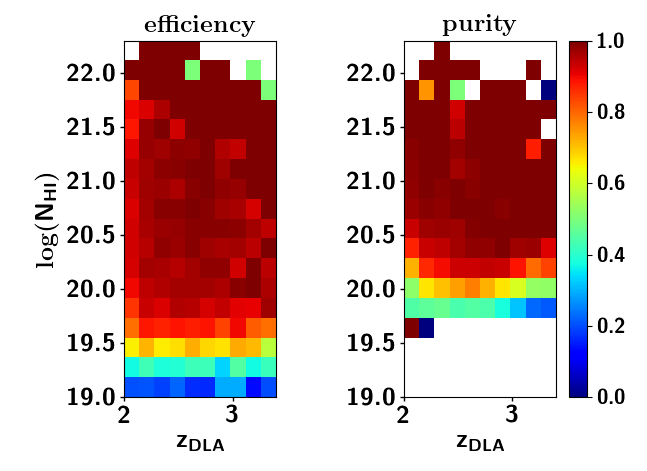}
   \caption{Same than for Fig.~\ref{fig:eff_pur_vs_n_z} for forests with $\meanflux > 2 \times\fluxunit$.
    }
         \label{fig:eff_pur_vs_n_z_bright}
\end{figure}

\begin{figure}
   \centering
   \includegraphics[width = \columnwidth]{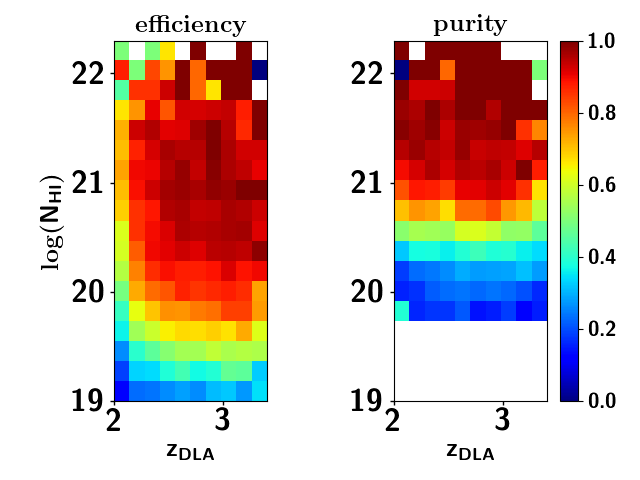}
   \caption{Same than for Fig.~\ref{fig:eff_pur_vs_n_z} for forests with $\meanflux < 2 \times\fluxunit$.
    }
         \label{fig:eff_pur_vs_n_z_faint}
\end{figure}

\begin{figure}
   \centering
   \includegraphics[width = \columnwidth]{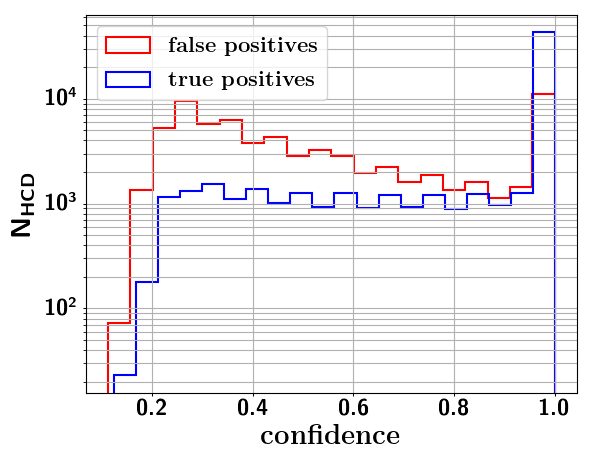}
   \caption{Confidence parameter distributions for the 132,226  absorbing systems found by the CNN: 69,380 false positives (red) and 62,846 true positives (blue). False positives are defined as HCDs found by the CNN but that do not matched HCD input. To match an input HCD, it must be on the same sightline with  $\Delta \zdla < 0.02$.
   }
         \label{fig:conf_hist}
\end{figure}

\subsection{Parameter estimation}
\label{sec:recovery}

\begin{figure}
   \centering
   \includegraphics[width = \columnwidth]{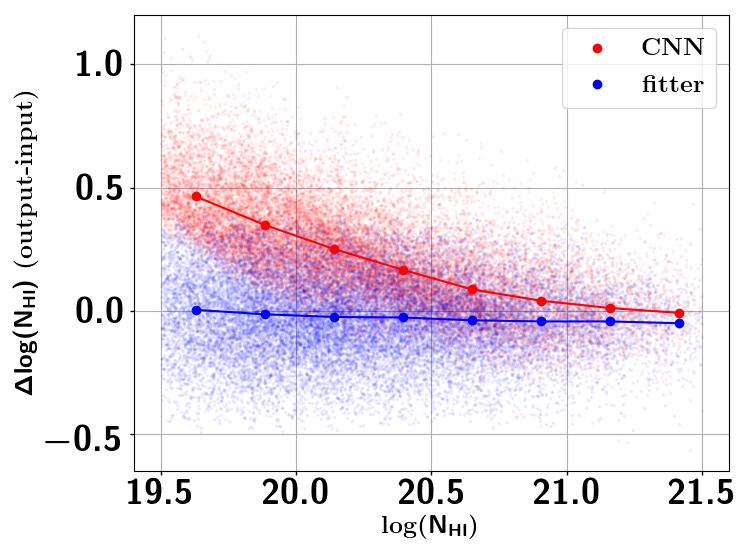}
   \caption{Difference in output and input values of $\nhi$  (in red for the CNN and blue for the fitter) vs input for the 21,234 found DLAs in the \lya~forest, i.e. in the rest-frame range  $1,040 \angstrom \leq \lambda_{\rm RF} \leq 1,216 \angstrom$.
   } 
         \label{fig:dnhi_vs_input}
\end{figure}

\begin{figure*}
   \centering
   \includegraphics[width=8cm]{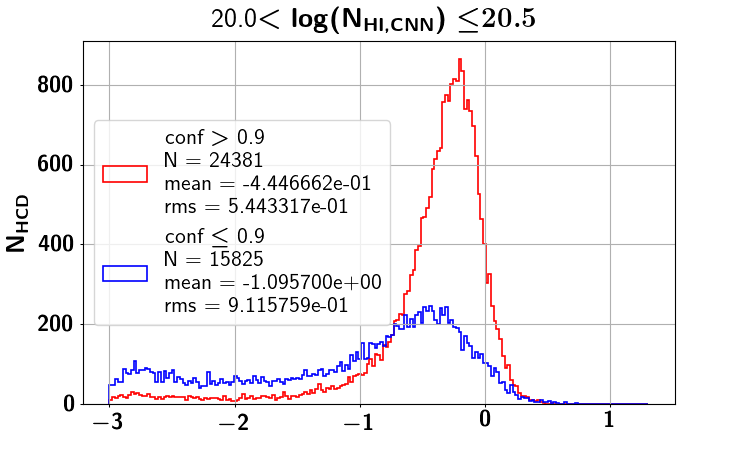}
   \includegraphics[width=8cm]{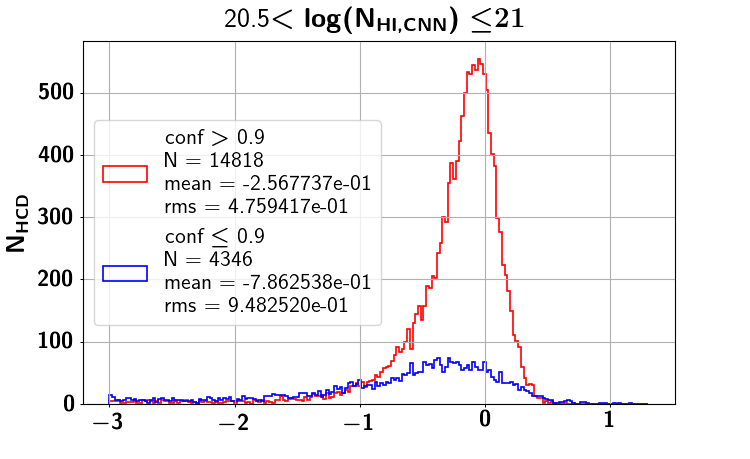}
   \includegraphics[width=8cm]{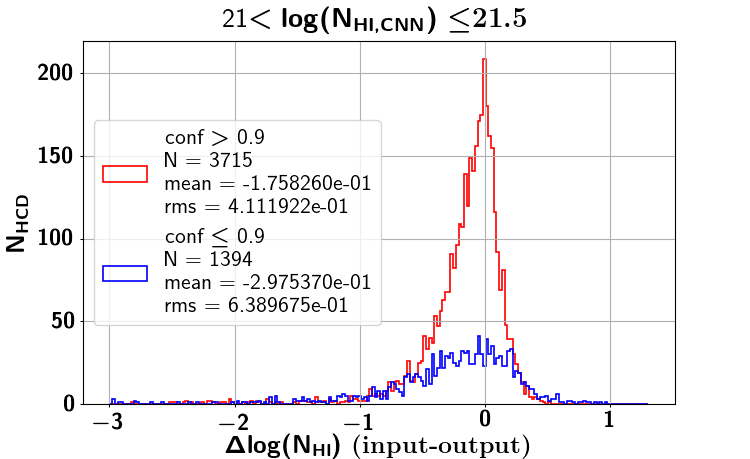}
   \includegraphics[width=8cm]{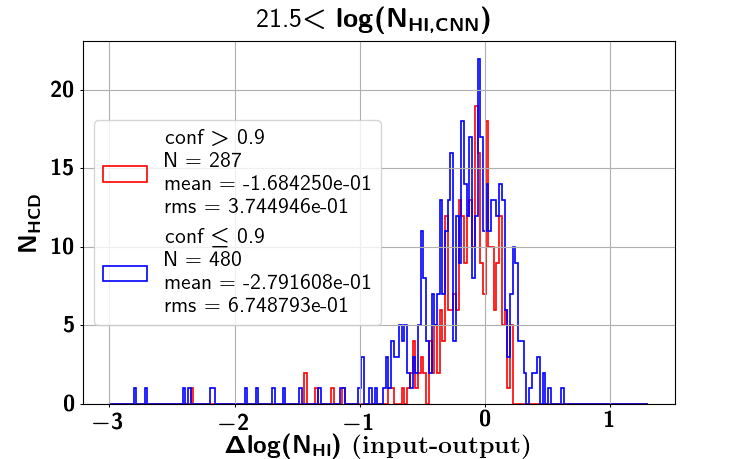}
   \caption{Distributions of the difference in $\lognhi$  between the CNN predictions and the mock inputs for DLA candidates  with confidence $> 0.9$ (red) and confidence $\leq 0.9$ (blue). The distributions are shown in four ranges of $\lognhi$ (CNN value) as labeled.
   }
         \label{fig:dnhi_conf}
\end{figure*}

  The CNN cannot be expected to give an unbiased estimate of $\lognhi$ because
  the DLA sample was selected by the CNN.
  The mocks contain a large number of low column-density HCDs
  (Fig.~\ref{fig:distrib_dla}) and some, through noise, may appear
  as detectable DLAs with $\lognhi>20$.  As such, we expect that
  the estimated $\lognhi$ to be on average greater than the true $\lognhi$.
This expectation is confirmed by
Fig.~\ref{fig:dnhi_vs_input} which compares the values of $\nhi$
returned by the finder with
the input value from the mocks.

We investigate the dependence of this systematic bias
on the confidence parameter in Fig.~\ref{fig:dnhi_conf} showing
the difference between input and CNN values of $\lognhi$ for
four ranges of CNN values of $\lognhi$.
First, as already seen in  Fig.~\ref{fig:dnhi_vs_input},
it shows that the bias is worse for low $\lognhi$ as the mean increases toward 0 when increasing $\lognhi$. But more importantly, it demonstrates that the confidence parameter is a good indicator of biased $\nhi$ since the blue curves always tend toward more negative values that the red curves.  The $\Delta \lognhi$ tail of non-confident HCDs is particularly long on the top left panel.
This is because even if these low-$\nhi$ candidates are matched to input HCDs,
we are close to the $\nhi$ detection threshold, so that many detected HCDs are in fact noise fluctuations close to a low-$\nhi$ HCD.  As such, the confidence parameter is also very useful to increase the purity in the low-$\nhi$ regime.

To provide a more unbiased estimate of $\lognhi$ we developed a
DLA fitter and applied it to DLA candidates
in the rest-frame range $1,040 \angstrom \leq \lambda_{\rm RF} \leq 1,216 \angstrom$ and $\log \nhi < 22$.
Fig.~\ref{fig:dnhi_vs_input} shows the difference between the input $\nhi$ and Voigt-profile fitted $\nhi$, which are more accurate that the CNN ones.

\section{The DR16 DLA catalog}
\label{sec:results}

We applied the automated algorithm to the 263,201 DR16 quasar spectra sample described in Sec.~\ref{sec:dr16q}. 
A total of 176,807 HCDs were found with $\zqso >$ $\zdla$ and $\zdla \geq 2$ in 112,155 sightlines.
These numbers reduced to 117,458 absorbers in 78,018 sightlines when we reject BAL quasars with {\sc BAL\_PROB} $> 0$, among them 39,067 (33\%) are classified as confident with confidence $> 0.9$.
Fig.~\ref{fig:dist_dr16} shows the $\zdla$ and $\lognhi$ distributions for the 20,375 bright forests and the remaining 97,083 faint forests of the 117,458 total sample.
 The sample was further reduced to 57,136 absorbers with $\log \nhi$ $\geq 20.3$ in 20,016 sightlines, yielding a purity of $\sim 0.3$ given that the number of DLAs per los should be roughly $< 1$~\citep{Noterdaeme2012, Bird2014}. Only considering bright forests raises the purity to be $>0.9$ for DLAs with $\log \nhi$ $\geq 20.3$ since the CNN found 6,996 such absorbers in 6,293 lines of sight.

The DLA sample we presented in DR16Q~\cite{Lyke2020} only include DLAs absorbers with $\lognhi \geq 20.3$ and confidence $>$ 0.9 and we did not reject BAL quasars. As presented in Sec.~\ref{sec:effpure}, the confidence cut highly degrades efficiency toward high $\nhi$ and low absorbing redshifts. The DLA sample presented here is consequently more complete and less pure.
As discussed in Sect.~\ref{sec:effpure}, users of this catalog can construct their own selection criteria, depending on their specific needs.

\begin{figure}
   \centering
      \includegraphics[width = 0.48 \columnwidth]{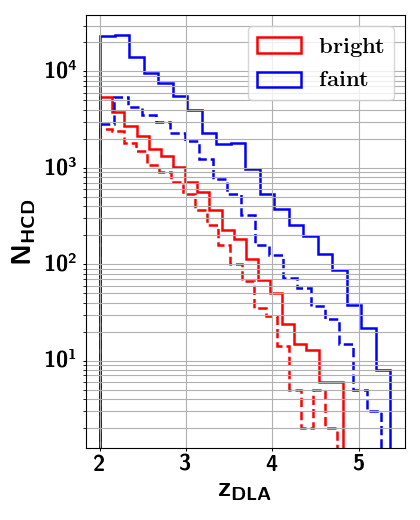}
   \includegraphics[width = 0.48 \columnwidth]{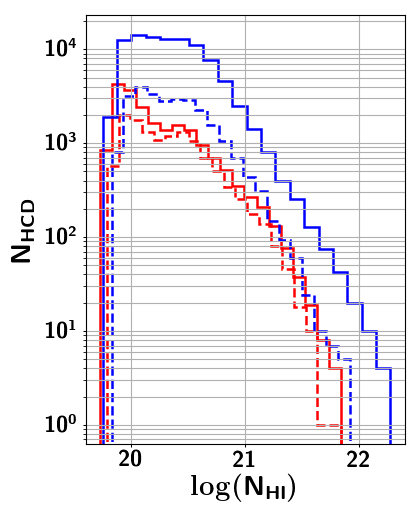}
   \caption{Distribution of $\zdla$ and $\lognhi$ of the 117,458 absorbers detected by the CNN in DR16Q for the 20,375 bright forests
     with $\meanflux>2\times\fluxunit$ (red) and the remaining 97,083 faint forests with $\meanflux \leq 2\times\fluxunit$ (blue). The dashed lines show the same samples reduced to confident absorbers only with confidence $\geq$ 0.9.}
         \label{fig:dist_dr16}
   \end{figure}

\begin{figure}
   \centering
   \includegraphics[width = \columnwidth]{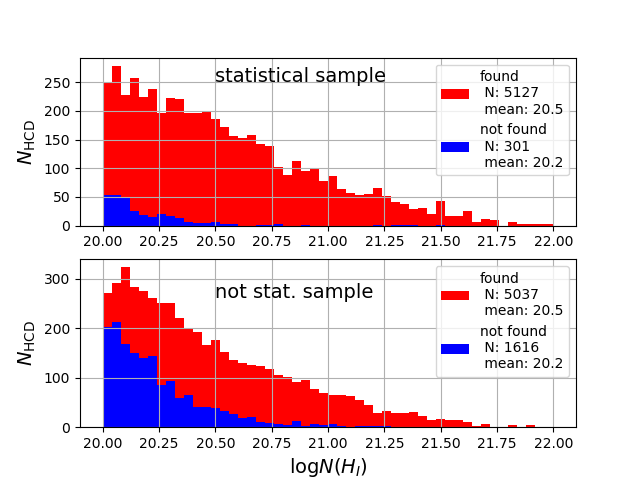}
   \caption{Distribution of $\nhi$ for the 12,081 DLAs of~\citet{Noterdaeme2012} (N12) found and not found by the CNN in red and blue respectively. The statistical sample consists of confident DLA candidates with high signal to noise which are used in~\cite{Noterdaeme2012} to measure the $\lognhi$ distribution and the cosmological mass density of neutral gas. 
    }
         \label{fig:nhi_found_notfound}
   \end{figure}
\begin{figure}
   \centering
   \includegraphics[width = \columnwidth]{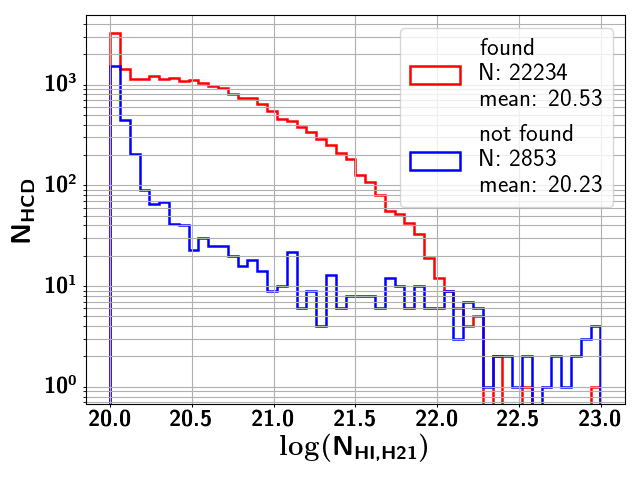}
   \caption{Distribution of $\nhi$ for the 25,087 confident DLAs
($p_{\rm DLA}>0.9$)
     of~\citet{Ho2021} (H21) found and not found by the CNN in red and blue respectively.
    }
         \label{fig:nhi_found_notfound_garnett}
   \end{figure}

\begin{figure}
   \centering
   \includegraphics[width = \columnwidth]{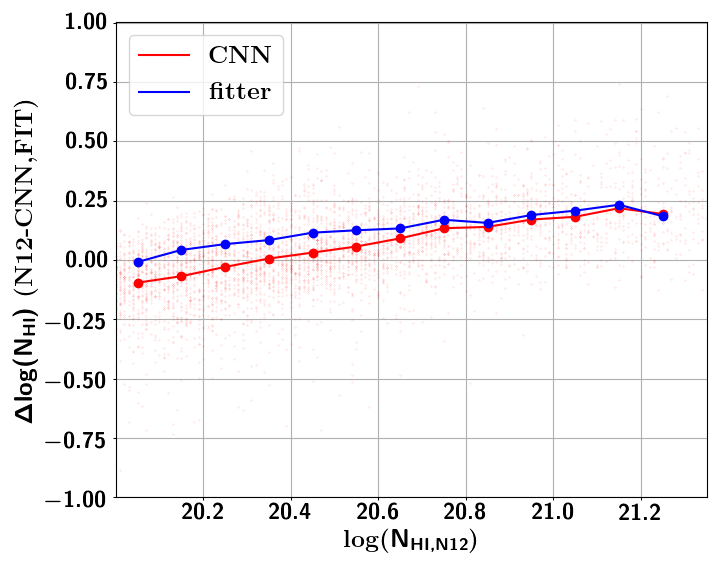}
   \caption{Comparison of the 4,620 $\nhi$ of the DR9 absorbers in the forest, i.e. in the rest-frame range  $1,040 \angstrom \leq \lambda_{\rm RF} \leq 1,216 \angstrom,$ found in both N12 and this study, using either the CNN (red) or the fitter (blue) result.    }
         \label{fig:nhidr9_nhidr16}
   \end{figure}

\begin{figure}
   \centering
\includegraphics[width = \columnwidth]{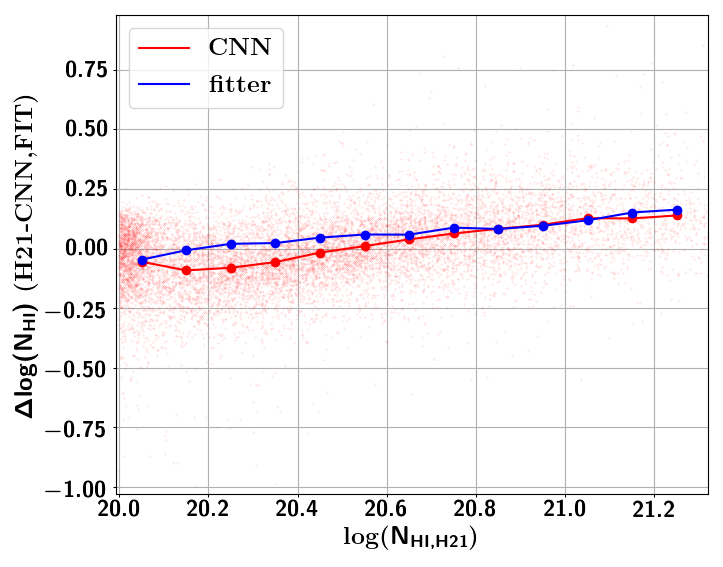}
   \caption{Same as Fig. \ref{fig:nhidr9_nhidr16} for H21. The sample contains 11,935 absorbers in the forest.
   }
         \label{fig:nhigarnett_nhidr16}
   \end{figure}

We compared our  sample of 117,458 absorbers with
two other catalogs based on BOSS and eBOSS data. 
The first is the 12,081 absorber sample of \citet{Noterdaeme2012},
hearafter N12, based
on the DR9 SDSS data release that uses Voigt profile fitting procedure.
The second, based on DR16 SDSS data, was provided by \citet{Ho2021},
hearafter H21, that extends the Gaussian Processes method presented in 
\citet{Garnett2017}. We reject BAL quasars with   {\sc BAL\_PROB} $> 0$ and
consider  only DLA with a high probability, $\rm p_{DLA} > 0.9$. With these criteria,
their sample contains a total of 25,087 absorbers.

Given that the efficiency and purity of the catalogs are functions
of cuts on signal-to-noise, mean forest flux, and the DLA parameters
$\lognhi$ and $\zdla$, we do not expect perfect overlap between the
catalogs.
This fact is illustrated in  Fig. \ref{fig:nhi_found_notfound}
for N12. As for the mock study in Sec.~\ref{sec:effpure}, DLAs are matched if they are in the same sightline and have absorbing redshifts such that $\rm \Delta \zdla < 0.02$
The figure shows their distribution of $\lognhi$ for candidate DLAs that
are found and not found in our catalog.  The distribution is
shown for the ``statistical'' and ``non-statistical'' samples of N12.
The statistical sample consists of confident DLA candidates 
with sufficiently high signal to noise to be used in~\cite{Noterdaeme2012} to
measure the $\lognhi$ distribution and the cosmological mass density of neutral gas.
We see that the overlap is very good for the statistical sample with
$\lognhi>20.2$.  On the other hand, the overlap for the non-statistical
sample is good only for $\lognhi>20.6$.

Figure \ref{fig:nhi_found_notfound_garnett} shows the same distribution
for the catalog of H21 where a similar behavior
is seen.

Figures~\ref{fig:nhidr9_nhidr16} and
\ref{fig:nhigarnett_nhidr16}
compare the values of $\lognhi$ from N12 and H21 with
our values as determined by the CNN and by our fitter.
The displayed samples are restricted to absorbers in the \lya~forest in order to have values of $\lognhi$ for the fitter as well. However, the trend of the difference with $\lognhi$ as predicted by the CNN (the blue curves) is similar when using the full sample of matched DLAs for both N12 and H21.
Agreement is found at the 0.1 level, with our values
being slightly greater than those of N12 and H21 at low $\nhi$
and slightly less than those of N12 and H21 at high $\nhi$.
We also note that the CNN finds values that are typically 0.1
greater than our fitter.

We make the catalog available \href{https://drive.google.com/drive/folders/1UaFHVwSNPpqkxTbcbR8mVRJ5BUR9KHzA?usp=sharing}{here} as a FITS file. There is a line for each detected absorber with $\zqso >\zdla$ and $\zdla \geq 2$ in sightline with  {\sc BAL\_PROB} $=0$. Each of the 117,458 line contains the following information:
\begin{itemize}
\item {\sc THING\_ID}: the SDSS identifier as found in DR16Q
\item {\sc Z\_QSO}: the quasar redshift of the sightline using  the {\sc Z\_LYAWG} estimator of DR16Q
\item {\sc PLATE}: SDSS spectroscopic plate of the sightline as found in DR16Q
\item {\sc MJD}: SDSS modified Julian date of observation  of the sightline as found in DR16Q
\item {\sc FIBERID}: SDSS spectroscopic fiber identification of the sightline as found in DR16Q
\item {\sc RA}: right ascencion of the sightline as found in DR16Q, in degrees
\item {\sc DEC}: declination of the sightline as found in DR16Q, in degrees
\item {\sc SNR}: mean signal-to-noise ratio of the sightline
\item {\sc MEAN\_FLUX}: mean forest flux in $\fluxunit$. The completeness and purity of sightlines with  {\sc MEAN\_FLUX} $ > 2 \times \fluxunit$ are greater than 90\% for absorbers with  $20.1\leq \lognhi \leq 22$. 
\item {\sc Z\_CNN}: the absorber redshift as found by the CNN
\item {\sc NHI\_CNN}: the logarithm of the absorber column density as found by the CNN
\item {\sc CONF\_CNN}: confidence parameter of the CNN over the range (0,1). Absorber with confidence $>0.5$ are considered as highly confident absorbers.
\item {\sc NHI\_FIT}: the logarithm of the absorber column density as found by the Voigt-profile fitter for absorbers in the rest-frame range $1,040 \angstrom \leq \lambda_{\rm RF} \leq 1,216 \angstrom$ and $\log \nhi < 22$. This parameter is set to -1 for absorbers that do not meet the criteria.
\end{itemize}

\section{Conclusions}
\label{sec:conclusion}

We presented here the production of the strong-absorber catalog in the 263,201 Ly$\alpha$ quasar spectra of the final SDSS-IV quasar catalog from DR16 \citep{Lyke2020}. We used the CNN pipeline from~\cite{Parks2018} to identify absorbers and estimate their properties $\zdla$ and $\nhi$. This choice was motivated by the fact that  the algorithm has been constructed for low redshift and low signal to noise BOSS/eBOSS quasar spectra.

We performed completeness and purity studies of the algorithm with synthetic spectra~\citep{Etourneau2021}
produced for the eBOSS Ly$\alpha$ data analysis~\citep{dMdB2020} that reproduce the characteristics of the data sample, in terms of redshift and signal-to-noise distribution.
The comparison between finder outputs and mock inputs showed that the algorithm performs well for confident DLAs with  $2.2 \leq \zdla \leq 3.5$, $20.3 \leq \lognhi \leq 21.5$ and confidence parameter $> 0.9$ with both purity and efficiency $>0.9$. Reducing the sample to bright forests only with $\meanflux> 2 \times\fluxunit$ increases efficiency and purity to  $>0.9$ values for a wider parameter range, for absorbers with $\lognhi \geq 20.1$.

We found a bias for $\nhi$ toward the lowest end because the finder detects absorbers with $\lognhi$ as low as 19 but overestimates this parameter just above the threshold it has been trained with. To alleviate this issue, we fit detected strong absorptions in the  rest-frame range $1,040 \angstrom \leq \lambda_{\rm RF} \leq 1,216 \angstrom$ with Voigt-profiles, which returns more accurate value of $\nhi$ than the CNN.

The algorithm detect 117,458 strong absorbers with $\lognhi > 19.7$ and 57,136 DLAs with $\lognhi > 20.3$, which is the largest DLA sample to date.   We provided the complete results of the finder for absorbers with $\zqso > \zdla$, $\zdla \geq 2$ and in sightlines without BALs detected in the DR16Q such that {\sc BAL\_PROB} $= 0$. We also provided $\nhi$ information of the Voigt-profile fitting for confident absorbers in the rest-frame range $1,040 \angstrom \leq \lambda_{\rm RF} \leq 1,216 \angstrom$. We compared our results to previously published catalogs from~\cite{Noterdaeme2012} and \cite{Ho2020} showing consistent findings.

This comprehensive analysis will enable users of this catalog to construct their own selection criteria matching the needs of their study. In addition, it highlights the regimes where DLA finders need to be improved, in particular the low-signal-to-noise regime. 

\acknowledgments

S.C. thanks Xavier Prochaska for his precious help and the many discussions for using the Convolutional Neural Network. The authors thank the DESI Ly$\alpha$ working group for providing the software to simulate the mock spectra. S.C. was partially supported by the DOE’s Office of Advanced Scientific Computing Research and Office of High Energy Physics through the Scientific Discovery through Advanced Computing (SciDAC) program.

This work was supported by the A*MIDEX project (ANR-11-IDEX-0001-02)funded by the “Investissements d'Avenir” French Government program, managed by the French National Research Agency (ANR), and by ANR under contracts ANR-14-ACHN-0021 and ANR-16-CE31–0021.

Funding for the Sloan Digital Sky Survey IV has been provided by the Alfred P. Sloan Foundation, the U.S. Department of Energy Office of Science, and the Participating Institutions. SDSS acknowledges support and resources from the Center for High-Performance Computing at the University of Utah. The SDSS web site is www.sdss.org. SDSS is managed by the Astrophysical Research Consortium for the Participating Institutions of the SDSS Collaboration including the Brazilian Participation Group, the Carnegie Institution for Science, Carnegie Mellon University, the Chilean Participation Group, the French Participation Group, Harvard-Smithsonian Center for Astrophysics, Instituto de Astrofísica de Canarias, The Johns Hopkins University, Kavli Institute for the Physics and Mathematics of the Universe (IPMU) / University of Tokyo, the Korean Participation Group, Lawrence Berkeley National Laboratory, Leibniz Institut f\"ur Astrophysik Potsdam (AIP), Max-Planck-Institut f\"ur Astronomie (MPIA Heidelberg), Max-Planck-Institut f\"ur Astrophysik (MPA Garching), Max-Planck-Institut f\"ur Extraterrestrische Physik (MPE), National Astronomical Observatories of China, New Mexico State University, New York University, University of Notre Dame, Observatório Nacional / MCTI, The Ohio State University, Pennsylvania State University, Shanghai Astronomical Observatory, United Kingdom Participation Group, Universidad Nacional Aut\'onoma de M\'exico, University of Arizona, University of Colorado Boulder, University of Oxford, University of Portsmouth, University of Utah, University of Virginia, University of Washington, University of Wisconsin, Vanderbilt University, and Yale University.

In addition, this research relied on resources provided to the eBOSS Collaboration by the National Energy Research Scientific Computing Center (NERSC).  NERSC is a U.S. Department of Energy Office of Science User Facility operated under Contract No. DE-AC02-05CH11231.

\vspace{5cm}


\bibliography{biblio}{}

\begin{thebibliography}{}
\expandafter\ifx\csname natexlab\endcsname\relax\def\natexlab#1{#1}\fi
\providecommand{\url}[1]{\href{#1}{#1}}
\providecommand{\dodoi}[1]{doi:~\href{http://doi.org/#1}{\nolinkurl{#1}}}
\providecommand{\doeprint}[1]{\href{http://ascl.net/#1}{\nolinkurl{http://ascl.net/#1}}}
\providecommand{\doarXiv}[1]{\href{https://arxiv.org/abs/#1}{\nolinkurl{https://arxiv.org/abs/#1}}}

\bibitem[{{Ahumada} {et~al.}(2020){Ahumada}, {Prieto}, {Almeida}, {Anders},
  {Anderson}, {Andrews}, {Anguiano}, {Arcodia}, {Armengaud}, {Aubert}, {Avila},
  {Avila-Reese}, {Badenes}, {Balland }, {Barger}, {Barrera-Ballesteros},
  {Basu}, {Bautista}, {Beaton}, {Beers}, {Benavides}, {Bender}, {Bernardi},
  {Bershady}, {Beutler}, {Bidin}, {Bird}, {Bizyaev}, {Blanc}, {Blanton},
  {Boquien}, {Borissova}, {Bovy}, {Brand t}, {Brinkmann}, {Brownstein},
  {Bundy}, {Bureau}, {Burgasser}, {Burtin}, {Cano-D{\'\i}az}, {Capasso},
  {Cappellari}, {Carrera}, {Chabanier}, {Chaplin}, {Chapman}, {Cherinka},
  {Chiappini}, {Doohyun Choi}, {Chojnowski}, {Chung}, {Clerc}, {Coffey},
  {Comerford}, {Comparat}, {da Costa}, {Cousinou}, {Covey}, {Crane}, {Cunha},
  {Ilha}, {Dai}, {Damsted}, {Darling}, {Davidson}, {Davies}, {Dawson}, {De},
  {de la Macorra}, {De Lee}, {Queiroz}, {Deconto Machado}, {de la Torre},
  {Dell'Agli}, {du Mas des Bourboux}, {Diamond-Stanic}, {Dillon}, {Donor},
  {Drory}, {Duckworth}, {Dwelly}, {Ebelke}, {Eftekharzadeh}, {Davis Eigenbrot},
  {Elsworth}, {Eracleous}, {Erfanianfar}, {Escoffier}, {Fan}, {Farr},
  {Fern{\'a}ndez-Trincado}, {Feuillet}, {Finoguenov}, {Fofie},
  {Fraser-McKelvie}, {Frinchaboy}, {Fromenteau}, {Fu}, {Galbany}, {Garcia},
  {Garc{\'\i}a-Hern{\'a}ndez}, {Oehmichen}, {Ge}, {Maia}, {Geisler}, {Gelfand
  }, {Goddy}, {Gonzalez-Perez}, {Grabowski}, {Green}, {Grier}, {Guo}, {Guy},
  {Harding}, {Hasselquist}, {Hawken}, {Hayes}, {Hearty}, {Hekker}, {Hogg},
  {Holtzman}, {Horta}, {Hou}, {Hsieh}, {Huber}, {Hunt}, {Chitham}, {Imig},
  {Jaber}, {Angel}, {Johnson}, {Jones}, {J{\"o}nsson}, {Jullo}, {Kim},
  {Kinemuchi}, {Kirkpatrick}, {Kite}, {Klaene}, {Kneib}, {Kollmeier}, {Kong},
  {Kounkel}, {Krishnarao}, {Lacerna}, {Lan}, {Lane}, {Law}, {Le Goff}, {Leung},
  {Lewis}, {Li}, {Lian}, {Lin}, {Long}, {Longa-Pe{\~n}a}, {Lundgren}, {Lyke},
  {Ted Mackereth}, {MacLeod}, {Majewski}, {Manchado}, {Maraston}, {Martini},
  {Masseron}, {Masters}, {Mathur}, {McDermid}, {Merloni}, {Merrifield},
  {M{\'e}sz{\'a}ros}, {Miglio}, {Minniti}, {Minsley}, {Miyaji}, {Mohammad},
  {Mosser}, {Mueller}, {Muna}, {Mu{\~n}oz-Guti{\'e}rrez}, {Myers}, {Nadathur},
  {Nair}, {Nandra}, {do Nascimento}, {Nevin}, {Newman}, {Nidever}, {Nitschelm},
  {Noterdaeme}, {O'Connell}, {Olmstead}, {Oravetz}, {Oravetz}, {Osorio},
  {Pace}, {Padilla}, {Palanque-Delabrouille}, {Palicio}, {Pan}, {Pan},
  {Parker}, {Paviot}, {Peirani}, {Ram{\'r}ez}, {Penny}, {Percival},
  {Perez-Fournon}, {P{\'e}rez-R{\`a}fols}, {Petitjean}, {Pieri},
  {Pinsonneault}, {Poovelil}, {Povick}, {Prakash}, {Price-Whelan}, {Raddick},
  {Raichoor}, {Ray}, {Rembold}, {Rezaie}, {Riffel}, {Riffel}, {Rix}, {Robin},
  {Roman-Lopes}, {Rom{\'a}n-Z{\'u}{\~n}iga}, {Rose}, {Ross}, {Rossi}, {Rowland
  s}, {Rubin}, {Salvato}, {S{\'a}nchez}, {S{\'a}nchez-Menguiano},
  {S{\'a}nchez-Gallego}, {Sayres}, {Schaefer}, {Schiavon}, {Schimoia},
  {Schlafly}, {Schlegel}, {Schneider}, {Schultheis}, {Schwope}, {Seo},
  {Serenelli}, {Shafieloo}, {Shamsi}, {Shao}, {Shen}, {Shetrone}, {Shirley},
  {Aguirre}, {Simon}, {Skrutskie}, {Slosar}, {Smethurst}, {Sobeck}, {Sodi},
  {Souto}, {Stark}, {Stassun}, {Steinmetz}, {Stello}, {Stermer},
  {Storchi-Bergmann}, {Streblyanska}, {Stringfellow}, {Stutz}, {Su{\'a}rez},
  {Sun}, {Taghizadeh-Popp}, {Talbot}, {Tayar}, {Thakar}, {Theriault}, {Thomas},
  {Thomas}, {Tinker}, {Tojeiro}, {Toledo}, {Tremonti}, {Troup}, {Tuttle},
  {Unda-Sanzana}, {Valentini}, {Vargas-Gonz{\'a}lez}, {Vargas-Maga{\~n}a},
  {V{\'a}zquez-Mata}, {Vivek}, {Wake}, {Wang}, {Weaver}, {Weijmans}, {Wild},
  {Wilson}, {Wilson}, {Wolthuis}, {Wood-Vasey}, {Yan}, {Yang}, {Y{\`e}che},
  {Zamora}, {Zarrouk}, {Zasowski}, {Zhang}, {Zhao}, {Zhao}, {Zheng}, {Zheng},
  {Zhu}, \& {Zou}}]{SDSSDR16}
{Ahumada}, R., {Prieto}, C.~A., {Almeida}, A., {et~al.} 2020, \apjs, 249, 3,
  \dodoi{10.3847/1538-4365/ab929e}

\bibitem[{{Bautista} {et~al.}(2017){Bautista}, {Busca}, {Guy}, {Rich},
  {Blomqvist}, {du Mas des Bourboux}, {Pieri}, {Font-Ribera}, {Bailey},
  {Delubac}, {Kirkby}, {Le Goff}, {Margala}, {Slosar}, {Vazquez}, {Brownstein},
  {Dawson}, {Eisenstein}, {Miralda-Escud{\'e}}, {Noterdaeme},
  {Palanque-Delabrouille}, {P{\^a}ris}, {Petitjean}, {Ross}, {Schneider},
  {Weinberg}, \& {Y{\`e}che}}]{Bautista2017}
{Bautista}, J.~E., {Busca}, N.~G., {Guy}, J., {et~al.} 2017, \aap, 603, A12,
  \dodoi{10.1051/0004-6361/201730533}

\bibitem[{{Bird} {et~al.}(2014){Bird}, {Vogelsberger}, {Haehnelt}, {Sijacki},
  {Genel}, {Torrey}, {Springel}, \& {Hernquist}}]{Bird2014}
{Bird}, S., {Vogelsberger}, M., {Haehnelt}, M., {et~al.} 2014, \mnras, 445,
  2313, \dodoi{10.1093/mnras/stu1923}

\bibitem[{{Blanton} {et~al.}(2017){Blanton}, {Bershady}, {Abolfathi},
  {Albareti}, {Allende Prieto}, {Almeida}, {Alonso-Garc{\'\i}a}, {Anders},
  {Anderson}, {Andrews}, {Aquino-Ort{\'\i}z}, {Arag{\'o}n-Salamanca},
  {Argudo-Fern{\'a}ndez}, {Armengaud}, {Aubourg}, {Avila-Reese}, {Badenes},
  {Bailey}, {Barger}, {Barrera-Ballesteros}, {Bartosz}, {Bates}, {Baumgarten},
  {Bautista}, {Beaton}, {Beers}, {Belfiore}, {Bender}, {Berlind}, {Bernardi},
  {Beutler}, {Bird}, {Bizyaev}, {Blanc}, {Blomqvist}, {Bolton}, {Boquien},
  {Borissova}, {van den Bosch}, {Bovy}, {Brandt}, {Brinkmann}, {Brownstein},
  {Bundy}, {Burgasser}, {Burtin}, {Busca}, {Cappellari}, {Delgado Carigi},
  {Carlberg}, {Carnero Rosell}, {Carrera}, {Chanover}, {Cherinka}, {Cheung},
  {G{\'o}mez Maqueo Chew}, {Chiappini}, {Choi}, {Chojnowski}, {Chuang},
  {Chung}, {Cirolini}, {Clerc}, {Cohen}, {Comparat}, {da Costa}, {Cousinou},
  {Covey}, {Crane}, {Croft}, {Cruz-Gonzalez}, {Garrido Cuadra}, {Cunha},
  {Damke}, {Darling}, {Davies}, {Dawson}, {de la Macorra}, {Dell'Agli}, {De
  Lee}, {Delubac}, {Di Mille}, {Diamond-Stanic}, {Cano-D{\'\i}az}, {Donor},
  {Downes}, {Drory}, {du Mas des Bourboux}, {Duckworth}, {Dwelly}, {Dyer},
  {Ebelke}, {Eigenbrot}, {Eisenstein}, {Emsellem}, {Eracleous}, {Escoffier},
  {Evans}, {Fan}, {Fern{\'a}ndez-Alvar}, {Fernandez-Trincado}, {Feuillet},
  {Finoguenov}, {Fleming}, {Font-Ribera}, {Fredrickson}, {Freischlad},
  {Frinchaboy}, {Fuentes}, {Galbany}, {Garcia-Dias},
  {Garc{\'\i}a-Hern{\'a}ndez}, {Gaulme}, {Geisler}, {Gelfand},
  {Gil-Mar{\'\i}n}, {Gillespie}, {Goddard}, {Gonzalez-Perez}, {Grabowski},
  {Green}, {Grier}, {Gunn}, {Guo}, {Guy}, {Hagen}, {Hahn}, {Hall}, {Harding},
  {Hasselquist}, {Hawley}, {Hearty}, {Gonzalez Hern{\'a}ndez}, {Ho}, {Hogg},
  {Holley-Bockelmann}, {Holtzman}, {Holzer}, {Huehnerhoff}, {Hutchinson},
  {Hwang}, {Ibarra-Medel}, {da Silva Ilha}, {Ivans}, {Ivory}, {Jackson},
  {Jensen}, {Johnson}, {Jones}, {J{\"o}nsson}, {Jullo}, {Kamble}, {Kinemuchi},
  {Kirkby}, {Kitaura}, {Klaene}, {Knapp}, {Kneib}, {Kollmeier}, {Lacerna},
  {Lane}, {Lang}, {Law}, {Lazarz}, {Lee}, {Le Goff}, {Liang}, {Li}, {Li},
  {Lian}, {Lima}, {Lin}, {Lin}, {Bertran de Lis}, {Liu}, {de Icaza Lizaola},
  {Long}, {Lucatello}, {Lundgren}, {MacDonald}, {Deconto Machado}, {MacLeod},
  {Mahadevan}, {Geimba Maia}, {Maiolino}, {Majewski}, {Malanushenko},
  {Malanushenko}, {Manchado}, {Mao}, {Maraston}, {Marques-Chaves}, {Masseron},
  {Masters}, {McBride}, {McDermid}, {McGrath}, {McGreer}, {Medina Pe{\~n}a},
  {Melendez}, {Merloni}, {Merrifield}, {Meszaros}, {Meza}, {Minchev},
  {Minniti}, {Miyaji}, {More}, {Mulchaey}, {M{\"u}ller-S{\'a}nchez}, {Muna},
  {Munoz}, {Myers}, {Nair}, {Nandra}, {Correa do Nascimento}, {Negrete},
  {Ness}, {Newman}, {Nichol}, {Nidever}, {Nitschelm}, {Ntelis}, {O'Connell},
  {Oelkers}, {Oravetz}, {Oravetz}, {Pace}, {Padilla}, {Palanque-Delabrouille},
  {Alonso Palicio}, {Pan}, {Parejko}, {Parikh}, {P{\^a}ris}, {Park}, {Patten},
  {Peirani}, {Pellejero-Ibanez}, {Penny}, {Percival}, {Perez-Fournon},
  {Petitjean}, {Pieri}, {Pinsonneault}, {Pisani}, {Poleski}, {Prada},
  {Prakash}, {Queiroz}, {Raddick}, {Raichoor}, {Barboza Rembold}, {Richstein},
  {Riffel}, {Riffel}, {Rix}, {Robin}, {Rockosi}, {Rodr{\'\i}guez-Torres},
  {Roman-Lopes}, {Rom{\'a}n-Z{\'u}{\~n}iga}, {Rosado}, {Ross}, {Rossi}, {Ruan},
  {Ruggeri}, {Rykoff}, {Salazar-Albornoz}, {Salvato}, {S{\'a}nchez}, {Aguado},
  {S{\'a}nchez-Gallego}, {Santana}, {Santiago}, {Sayres}, {Schiavon}, {da Silva
  Schimoia}, {Schlafly}, {Schlegel}, {Schneider}, {Schultheis}, {Schuster},
  {Schwope}, {Seo}, {Shao}, {Shen}, {Shetrone}, {Shull}, {Simon}, {Skinner},
  {Skrutskie}, {Slosar}, {Smith}, {Sobeck}, {Sobreira}, {Somers}, {Souto},
  {Stark}, {Stassun}, {Stauffer}, {Steinmetz}, {Storchi-Bergmann},
  {Streblyanska}, {Stringfellow}, {Su{\'a}rez}, {Sun}, {Suzuki}, {Szigeti},
  {Taghizadeh-Popp}, {Tang}, {Tao}, {Tayar}, {Tembe}, {Teske}, {Thakar},
  {Thomas}, {Thompson}, {Tinker}, {Tissera}, {Tojeiro}, {Hernandez Toledo}, {de
  la Torre}, {Tremonti}, {Troup}, {Valenzuela}, {Martinez Valpuesta},
  {Vargas-Gonz{\'a}lez}, {Vargas-Maga{\~n}a}, {Vazquez}, {Villanova}, {Vivek},
  {Vogt}, {Wake}, {Walterbos}, {Wang}, {Weaver}, {Weijmans}, {Weinberg},
  {Westfall}, {Whelan}, {Wild}, {Wilson}, {Wood-Vasey}, {Wylezalek}, {Xiao},
  {Yan}, {Yang}, {Ybarra}, {Y{\`e}che}, {Zakamska}, {Zamora}, {Zarrouk},
  {Zasowski}, {Zhang}, {Zhao}, {Zheng}, {Zheng}, {Zhou}, {Zhou}, {Zhu},
  {Zoccali}, \& {Zou}}]{Blanton2017}
{Blanton}, M.~R., {Bershady}, M.~A., {Abolfathi}, B., {et~al.} 2017, \aj, 154,
  28, \dodoi{10.3847/1538-3881/aa7567}

\bibitem[{{Cen}(2012)}]{Cen2012}
{Cen}, R. 2012, \apj, 748, 121, \dodoi{10.1088/0004-637X/748/2/121}

\bibitem[{{Chabanier} {et~al.}(2019){Chabanier}, {Palanque-Delabrouille},
  {Y{\`e}che}, {Le Goff}, {Armengaud}, {Bautista}, {Blomqvist}, {Busca},
  {Dawson}, {Etourneau}, {Font-Ribera}, {Lee}, {du Mas des Bourboux}, {Pieri},
  {Rich}, {Rossi}, {Schneider}, \& {Slosar}}]{Chabanier2019}
{Chabanier}, S., {Palanque-Delabrouille}, N., {Y{\`e}che}, C., {et~al.} 2019,
  \jcap, 2019, 017, \dodoi{10.1088/1475-7516/2019/07/017}

\bibitem[{{Dawson} {et~al.}(2016){Dawson}, {Kneib}, {Percival}, {Alam},
  {Albareti}, {Anderson}, {Armengaud}, {Aubourg}, {Bailey}, {Bautista},
  {Berlind}, {Bershady}, {Beutler}, {Bizyaev}, {Blanton}, {Blomqvist},
  {Bolton}, {Bovy}, {Brandt}, {Brinkmann}, {Brownstein}, {Burtin}, {Busca},
  {Cai}, {Chuang}, {Clerc}, {Comparat}, {Cope}, {Croft}, {Cruz-Gonzalez}, {da
  Costa}, {Cousinou}, {Darling}, {de la Macorra}, {de la Torre}, {Delubac}, {du
  Mas des Bourboux}, {Dwelly}, {Ealet}, {Eisenstein}, {Eracleous}, {Escoffier},
  {Fan}, {Finoguenov}, {Font-Ribera}, {Frinchaboy}, {Gaulme}, {Georgakakis},
  {Green}, {Guo}, {Guy}, {Ho}, {Holder}, {Huehnerhoff}, {Hutchinson}, {Jing},
  {Jullo}, {Kamble}, {Kinemuchi}, {Kirkby}, {Kitaura}, {Klaene}, {Laher},
  {Lang}, {Laurent}, {Le Goff}, {Li}, {Liang}, {Lima}, {Lin}, {Lin}, {Lin},
  {Long}, {Lundgren}, {MacDonald}, {Geimba Maia}, {Malanushenko},
  {Malanushenko}, {Mariappan}, {McBride}, {McGreer}, {M{\'e}nard}, {Merloni},
  {Meza}, {Montero-Dorta}, {Muna}, {Myers}, {Nandra}, {Naugle}, {Newman},
  {Noterdaeme}, {Nugent}, {Ogando}, {Olmstead}, {Oravetz}, {Oravetz},
  {Padmanabhan}, {Palanque-Delabrouille}, {Pan}, {Parejko}, {P{\^a}ris},
  {Peacock}, {Petitjean}, {Pieri}, {Pisani}, {Prada}, {Prakash}, {Raichoor},
  {Reid}, {Rich}, {Ridl}, {Rodriguez-Torres}, {Carnero Rosell}, {Ross},
  {Rossi}, {Ruan}, {Salvato}, {Sayres}, {Schneider}, {Schlegel}, {Seljak},
  {Seo}, {Sesar}, {Shandera}, {Shu}, {Slosar}, {Sobreira}, {Streblyanska},
  {Suzuki}, {Taylor}, {Tao}, {Tinker}, {Tojeiro}, {Vargas-Maga{\~n}a}, {Wang},
  {Weaver}, {Weinberg}, {White}, {Wood-Vasey}, {Yeche}, {Zhai}, {Zhao}, {Zhao},
  {Zheng}, {Ben Zhu}, \& {Zou}}]{Dawson2016}
{Dawson}, K.~S., {Kneib}, J.-P., {Percival}, W.~J., {et~al.} 2016, \aj, 151,
  44, \dodoi{10.3847/0004-6256/151/2/44}

\bibitem[{{du Mas des Bourboux} {et~al.}(2020){du Mas des Bourboux}, {Rich},
  {Font-Ribera}, {de Sainte Agathe}, {Farr}, {Etourneau}, {Le Goff}, {Cuceu},
  {Balland}, {Bautista}, {Blomqvist}, {Brinkmann}, {Brownstein}, {Chabanier},
  {Chaussidon}, {Dawson}, {Gonz{\'a}lez-Morales}, {Guy}, {Lyke}, {de la
  Macorra}, {Mueller}, {Myers}, {Nitschelm}, {Mu{\~n}oz Guti{\'e}rrez},
  {Palanque-Delabrouille}, {Parker}, {Percival}, {P{\'e}rez-R{\`a}fols},
  {Petitjean}, {Pieri}, {Ravoux}, {Rossi}, {Schneider}, {Seo}, {Slosar},
  {Stermer}, {Vivek}, {Y{\`e}che}, \& {Youles}}]{dMdB2020}
{du Mas des Bourboux}, H., {Rich}, J., {Font-Ribera}, A., {et~al.} 2020, \apj,
  901, 153, \dodoi{10.3847/1538-4357/abb085}

\bibitem[{{Etourneau} {et~al.}(2021){Etourneau}, {Le Goff}, {Busca}, Julianna,
  Jim, bla, \& bla}]{Etourneau2021}
{Etourneau}, T., {Le Goff}, J.-M., {Busca}, N., {et~al.} 2021, to be published
  in \jcap

\bibitem[{{Font-Ribera} \& {Miralda-Escud{\'e}}(2012)}]{FontRibera2012}
{Font-Ribera}, A., \& {Miralda-Escud{\'e}}, J. 2012, \jcap, 2012, 028,
  \dodoi{10.1088/1475-7516/2012/07/028}

\bibitem[{{Fumagalli} {et~al.}(2020){Fumagalli}, {Fotopoulou}, \&
  {Thomson}}]{Fumagalli2020}
{Fumagalli}, M., {Fotopoulou}, S., \& {Thomson}, L. 2020, \mnras, 498, 1951,
  \dodoi{10.1093/mnras/staa2388}

\bibitem[{{Fumagalli} {et~al.}(2016){Fumagalli}, {O'Meara}, \&
  {Prochaska}}]{Fumagalli2016}
{Fumagalli}, M., {O'Meara}, J.~M., \& {Prochaska}, J.~X. 2016, \mnras, 455,
  4100, \dodoi{10.1093/mnras/stv2616}

\bibitem[{{Fumagalli} {et~al.}(2014){Fumagalli}, {O'Meara}, {Prochaska},
  {Kanekar}, \& {Wolfe}}]{Fumagalli2014}
{Fumagalli}, M., {O'Meara}, J.~M., {Prochaska}, J.~X., {Kanekar}, N., \&
  {Wolfe}, A.~M. 2014, \mnras, 444, 1282, \dodoi{10.1093/mnras/stu1512}

\bibitem[{{Fumagalli} {et~al.}(2013){Fumagalli}, {O'Meara}, {Prochaska}, \&
  {Worseck}}]{Fumagalli2013}
{Fumagalli}, M., {O'Meara}, J.~M., {Prochaska}, J.~X., \& {Worseck}, G. 2013,
  \apj, 775, 78, \dodoi{10.1088/0004-637X/775/1/78}

\bibitem[{{Gardner} {et~al.}(1997){Gardner}, {Katz}, {Hernquist}, \&
  {Weinberg}}]{Gardner1997}
{Gardner}, J.~P., {Katz}, N., {Hernquist}, L., \& {Weinberg}, D.~H. 1997, \apj,
  484, 31, \dodoi{10.1086/304310}

\bibitem[{{Garnett} {et~al.}(2017){Garnett}, {Ho}, {Bird}, \&
  {Schneider}}]{Garnett2017}
{Garnett}, R., {Ho}, S., {Bird}, S., \& {Schneider}, J. 2017, \mnras, 472,
  1850, \dodoi{10.1093/mnras/stx1958}

\bibitem[{{Gunn} {et~al.}(2006){Gunn}, {Siegmund}, {Mannery}, {Owen}, {Hull},
  {Leger}, {Carey}, {Knapp}, {York}, {Boroski}, {Kent}, {Lupton}, {Rockosi},
  {Evans}, {Waddell}, {Anderson}, {Annis}, {Barentine}, {Bartoszek}, {Bastian},
  {Bracker}, {Brewington}, {Briegel}, {Brinkmann}, {Brown}, {Carr},
  {Czarapata}, {Drennan}, {Dombeck}, {Federwitz}, {Gillespie}, {Gonzales},
  {Hansen}, {Harvanek}, {Hayes}, {Jordan}, {Kinney}, {Klaene}, {Kleinman},
  {Kron}, {Kresinski}, {Lee}, {Limmongkol}, {Lindenmeyer}, {Long}, {Loomis},
  {McGehee}, {Mantsch}, {Neilsen}, {Neswold}, {Newman}, {Nitta}, {Peoples},
  {Pier}, {Prieto}, {Prosapio}, {Rivetta}, {Schneider}, {Snedden}, \&
  {Wang}}]{Gunn2006}
{Gunn}, J.~E., {Siegmund}, W.~A., {Mannery}, E.~J., {et~al.} 2006, \aj, 131,
  2332, \dodoi{10.1086/500975}

\bibitem[{{Haehnelt} {et~al.}(1998){Haehnelt}, {Steinmetz}, \&
  {Rauch}}]{Haehnelt1998}
{Haehnelt}, M.~G., {Steinmetz}, M., \& {Rauch}, M. 1998, \apj, 495, 647,
  \dodoi{10.1086/305323}

\bibitem[{{Ho} {et~al.}(2020){Ho}, {Bird}, \& {Garnett}}]{Ho2020}
{Ho}, M.-F., {Bird}, S., \& {Garnett}, R. 2020, \mnras, 496, 5436,
  \dodoi{10.1093/mnras/staa1806}

\bibitem[{{Ho} {et~al.}(2021){Ho}, {Bird}, \& {Garnett}}]{Ho2021}
---. 2021, arXiv e-prints, arXiv:2103.10964.
\newblock \doarXiv{2103.10964}

\bibitem[{Kingma \& Ba(2014)}]{Adam2014}
Kingma, D.~P., \& Ba, J. 2014, Adam: A Method for Stochastic Optimization.
\newblock \doarXiv{1412.6980}

\bibitem[{{Lee} {et~al.}(2015){Lee}, {Hennawi}, {Spergel}, {Weinberg}, {Hogg},
  {Viel}, {Bolton}, {Bailey}, {Pieri}, {Carithers}, {Schlegel}, {Lundgren},
  {Palanque-Delabrouille}, {Suzuki}, {Schneider}, \& {Y{\`e}che}}]{Lee2015}
{Lee}, K.-G., {Hennawi}, J.~F., {Spergel}, D.~N., {et~al.} 2015, \apj, 799,
  196, \dodoi{10.1088/0004-637X/799/2/196}

\bibitem[{{Lyke} {et~al.}(2020){Lyke}, {Higley}, {McLane}, {Schurhammer},
  {Myers}, {Ross}, {Dawson}, {Chabanier}, {Martini}, {Busca}, {du Mas des
  Bourboux}, {Salvato}, {Streblyanska}, {Zarrouk}, {Burtin}, {Anderson},
  {Bautista}, {Bizyaev}, {Brandt}, {Brinkmann}, {Brownstein}, {Comparat},
  {Green}, {de la Macorra}, {Mu{\~n}oz Guti{\'e}rrez}, {Hou}, {Newman},
  {Palanque-Delabrouille}, {P{\^a}ris}, {Percival}, {Petitjean}, {Rich},
  {Rossi}, {Schneider}, {Smith}, {Vivek}, \& {Weaver}}]{Lyke2020}
{Lyke}, B.~W., {Higley}, A.~N., {McLane}, J.~N., {et~al.} 2020, arXiv e-prints,
  arXiv:2007.09001.
\newblock \doarXiv{2007.09001}

\bibitem[{{McDonald} {et~al.}(2006){McDonald}, {Seljak}, {Burles}, {Schlegel},
  {Weinberg}, {Cen}, {Shih}, {Schaye}, {Schneider}, {Bahcall}, {Briggs},
  {Brinkmann}, {Brunner}, {Fukugita}, {Gunn}, {Ivezi{\'c}}, {Kent}, {Lupton},
  \& {Vanden Berk}}]{McDonald2006}
{McDonald}, P., {Seljak}, U., {Burles}, S., {et~al.} 2006, \apjs, 163, 80,
  \dodoi{10.1086/444361}

\bibitem[{{Myers} {et~al.}(2015){Myers}, {Palanque-Delabrouille}, {Prakash},
  {P{\^a}ris}, {Yeche}, {Dawson}, {Bovy}, {Lang}, {Schlegel}, {Newman},
  {Petitjean}, {Kneib}, {Laurent}, {Percival}, {Ross}, {Seo}, {Tinker},
  {Armengaud}, {Brownstein}, {Burtin}, {Cai}, {Comparat}, {Kasliwal},
  {Kulkarni}, {Laher}, {Levitan}, {McBride}, {McGreer}, {Miller}, {Nugent},
  {Ofek}, {Rossi}, {Ruan}, {Schneider}, {Sesar}, {Streblyanska}, \&
  {Surace}}]{Myers2015}
{Myers}, A.~D., {Palanque-Delabrouille}, N., {Prakash}, A., {et~al.} 2015,
  \apjs, 221, 27, \dodoi{10.1088/0067-0049/221/2/27}

\bibitem[{{Noterdaeme} {et~al.}(2009){Noterdaeme}, {Petitjean}, {Ledoux}, \&
  {Srianand}}]{Noterdaeme2009}
{Noterdaeme}, P., {Petitjean}, P., {Ledoux}, C., \& {Srianand}, R. 2009, \aap,
  505, 1087, \dodoi{10.1051/0004-6361/200912768}

\bibitem[{{Noterdaeme} {et~al.}(2012){Noterdaeme}, {Petitjean}, {Carithers},
  {P{\^a}ris}, {Font-Ribera}, {Bailey}, {Aubourg}, {Bizyaev}, {Ebelke},
  {Finley}, {Ge}, {Malanushenko}, {Malanushenko}, {Miralda-Escud{\'e}},
  {Myers}, {Oravetz}, {Pan}, {Pieri}, {Ross}, {Schneider}, {Simmons}, \&
  {York}}]{Noterdaeme2012}
{Noterdaeme}, P., {Petitjean}, P., {Carithers}, W.~C., {et~al.} 2012, \aap,
  547, L1, \dodoi{10.1051/0004-6361/201220259}

\bibitem[{{Ota} {et~al.}(2014){Ota}, {Walter}, {Ohta}, {Hatsukade}, {Carilli},
  {da Cunha}, {Gonz{\'a}lez-L{\'o}pez}, {Decarli}, {Hodge}, {Nagai}, {Egami},
  {Jiang}, {Iye}, {Kashikawa}, {Riechers}, {Bertoldi}, {Cox}, {Neri}, \&
  {Weiss}}]{Ota2014}
{Ota}, K., {Walter}, F., {Ohta}, K., {et~al.} 2014, \apj, 792, 34,
  \dodoi{10.1088/0004-637X/792/1/34}

\bibitem[{{Palanque-Delabrouille} {et~al.}(2013){Palanque-Delabrouille},
  {Y{\`e}che}, {Borde}, {Le Goff}, {Rossi}, {Viel}, {Aubourg}, {Bailey},
  {Bautista}, {Blomqvist}, {Bolton}, {Bolton}, {Busca}, {Carithers}, {Croft},
  {Dawson}, {Delubac}, {Font-Ribera}, {Ho}, {Kirkby}, {Lee}, {Margala},
  {Miralda-Escud{\'e}}, {Muna}, {Myers}, {Noterdaeme}, {P{\^a}ris},
  {Petitjean}, {Pieri}, {Rich}, {Rollinde}, {Ross}, {Schlegel}, {Schneider},
  {Slosar}, \& {Weinberg}}]{PalanqueDelabrouille2013}
{Palanque-Delabrouille}, N., {Y{\`e}che}, C., {Borde}, A., {et~al.} 2013, \aap,
  559, A85, \dodoi{10.1051/0004-6361/201322130}

\bibitem[{{Parks} {et~al.}(2018){Parks}, {Prochaska}, {Dong}, \&
  {Cai}}]{Parks2018}
{Parks}, D., {Prochaska}, J.~X., {Dong}, S., \& {Cai}, Z. 2018, \mnras, 476,
  1151, \dodoi{10.1093/mnras/sty196}

\bibitem[{{Petitjean} {et~al.}(2000){Petitjean}, {Srianand}, \&
  {Ledoux}}]{Petitjean2000}
{Petitjean}, P., {Srianand}, R., \& {Ledoux}, C. 2000, \aap, 364, L26.
\newblock \doarXiv{astro-ph/0011437}

\bibitem[{{Pontzen} {et~al.}(2008){Pontzen}, {Governato}, {Pettini}, {Booth},
  {Stinson}, {Wadsley}, {Brooks}, {Quinn}, \& {Haehnelt}}]{Pontzen2008}
{Pontzen}, A., {Governato}, F., {Pettini}, M., {et~al.} 2008, \mnras, 390,
  1349, \dodoi{10.1111/j.1365-2966.2008.13782.x}

\bibitem[{{Prochaska} {et~al.}(2005){Prochaska}, {Herbert-Fort}, \&
  {Wolfe}}]{Prochaska2005}
{Prochaska}, J.~X., {Herbert-Fort}, S., \& {Wolfe}, A.~M. 2005, \apj, 635, 123,
  \dodoi{10.1086/497287}

\bibitem[{{Prochaska} \& {Wolfe}(1997)}]{Prochaska1997}
{Prochaska}, J.~X., \& {Wolfe}, A.~M. 1997, \apj, 487, 73,
  \dodoi{10.1086/304591}

\bibitem[{{Prochaska} \& {Wolfe}(2009)}]{Prochaska2009}
---. 2009, \apj, 696, 1543, \dodoi{10.1088/0004-637X/696/2/1543}

\bibitem[{{Ross} {et~al.}(2012){Ross}, {Myers}, {Sheldon}, {Y{\`e}che},
  {Strauss}, {Bovy}, {Kirkpatrick}, {Richards}, {Aubourg}, {Blanton}, {Brandt},
  {Carithers}, {Croft}, {da Silva}, {Dawson}, {Eisenstein}, {Hennawi}, {Ho},
  {Hogg}, {Lee}, {Lundgren}, {McMahon}, {Miralda-Escud{\'e}},
  {Palanque-Delabrouille}, {P{\^a}ris}, {Petitjean}, {Pieri}, {Rich}, {Roe},
  {Schiminovich}, {Schlegel}, {Schneider}, {Slosar}, {Suzuki}, {Tinker},
  {Weinberg}, {Weyant}, {White}, \& {Wood-Vasey}}]{Ross2012}
{Ross}, N.~P., {Myers}, A.~D., {Sheldon}, E.~S., {et~al.} 2012, \apjs, 199, 3,
  \dodoi{10.1088/0067-0049/199/1/3}

\bibitem[{{Rudie} {et~al.}(2017){Rudie}, {Newman}, \& {Murphy}}]{Rudie2017}
{Rudie}, G.~C., {Newman}, A.~B., \& {Murphy}, M.~T. 2017, \apj, 843, 98,
  \dodoi{10.3847/1538-4357/aa74d7}

\bibitem[{{Slosar} {et~al.}(2011){Slosar}, {Font-Ribera}, {Pieri}, {Rich}, {Le
  Goff}, {Aubourg}, {Brinkmann}, {Busca}, {Carithers}, {Charlassier},
  {Cort{\^e}s}, {Croft}, {Dawson}, {Eisenstein}, {Hamilton}, {Ho}, {Lee},
  {Lupton}, {McDonald}, {Medolin}, {Muna}, {Miralda-Escud{\'e}}, {Myers},
  {Nichol}, {Palanque-Delabrouille}, {P{\^a}ris}, {Petitjean}, {Pi{\v{s}}kur},
  {Rollinde}, {Ross}, {Schlegel}, {Schneider}, {Sheldon}, {Weaver}, {Weinberg},
  {Yeche}, \& {York}}]{Slosar2011}
{Slosar}, A., {Font-Ribera}, A., {Pieri}, M.~M., {et~al.} 2011, \jcap, 2011,
  001, \dodoi{10.1088/1475-7516/2011/09/001}

\bibitem[{{Smee} {et~al.}(2013){Smee}, {Gunn}, {Uomoto}, {Roe}, {Schlegel},
  {Rockosi}, {Carr}, {Leger}, {Dawson}, {Olmstead}, {Brinkmann}, {Owen},
  {Barkhouser}, {Honscheid}, {Harding}, {Long}, {Lupton}, {Loomis}, {Anderson},
  {Annis}, {Bernardi}, {Bhardwaj}, {Bizyaev}, {Bolton}, {Brewington}, {Briggs},
  {Burles}, {Burns}, {Castander}, {Connolly}, {Davenport}, {Ebelke}, {Epps},
  {Feldman}, {Friedman}, {Frieman}, {Heckman}, {Hull}, {Knapp}, {Lawrence},
  {Loveday}, {Mannery}, {Malanushenko}, {Malanushenko}, {Merrelli}, {Muna},
  {Newman}, {Nichol}, {Oravetz}, {Pan}, {Pope}, {Ricketts}, {Shelden},
  {Sandford}, {Siegmund}, {Simmons}, {Smith}, {Snedden}, {Schneider},
  {SubbaRao}, {Tremonti}, {Waddell}, \& {York}}]{Smee2013}
{Smee}, S.~A., {Gunn}, J.~E., {Uomoto}, A., {et~al.} 2013, \aj, 146, 32,
  \dodoi{10.1088/0004-6256/146/2/32}

\bibitem[{{Vladilo} {et~al.}(2001){Vladilo}, {Centuri{\'o}n}, {Bonifacio}, \&
  {Howk}}]{Vladilo2001}
{Vladilo}, G., {Centuri{\'o}n}, M., {Bonifacio}, P., \& {Howk}, J.~C. 2001,
  \apj, 557, 1007, \dodoi{10.1086/321650}

\bibitem[{{Wolfe} {et~al.}(2005){Wolfe}, {Gawiser}, \& {Prochaska}}]{Wolfe2005}
{Wolfe}, A.~M., {Gawiser}, E., \& {Prochaska}, J.~X. 2005, \araa, 43, 861,
  \dodoi{10.1146/annurev.astro.42.053102.133950}

\bibitem[{{Wolfe} {et~al.}(1986){Wolfe}, {Turnshek}, {Smith}, \&
  {Cohen}}]{Wolfe1986}
{Wolfe}, A.~M., {Turnshek}, D.~A., {Smith}, H.~E., \& {Cohen}, R.~D. 1986,
  \apjs, 61, 249, \dodoi{10.1086/191114}

\bibitem[{{York} {et~al.}(2000){York}, {Adelman}, {Anderson}, {Anderson},
  {Annis}, {Bahcall}, {Bakken}, {Barkhouser}, {Bastian}, {Berman}, {Boroski},
  {Bracker}, {Briegel}, {Briggs}, {Brinkmann}, {Brunner}, {Burles}, {Carey},
  {Carr}, {Castander}, {Chen}, {Colestock}, {Connolly}, {Crocker}, {Csabai},
  {Czarapata}, {Davis}, {Doi}, {Dombeck}, {Eisenstein}, {Ellman}, {Elms},
  {Evans}, {Fan}, {Federwitz}, {Fiscelli}, {Friedman}, {Frieman}, {Fukugita},
  {Gillespie}, {Gunn}, {Gurbani}, {de Haas}, {Haldeman}, {Harris}, {Hayes},
  {Heckman}, {Hennessy}, {Hindsley}, {Holm}, {Holmgren}, {Huang}, {Hull},
  {Husby}, {Ichikawa}, {Ichikawa}, {Ivezi{\'c}}, {Kent}, {Kim}, {Kinney},
  {Klaene}, {Kleinman}, {Kleinman}, {Knapp}, {Korienek}, {Kron}, {Kunszt},
  {Lamb}, {Lee}, {Leger}, {Limmongkol}, {Lindenmeyer}, {Long}, {Loomis},
  {Loveday}, {Lucinio}, {Lupton}, {MacKinnon}, {Mannery}, {Mantsch}, {Margon},
  {McGehee}, {McKay}, {Meiksin}, {Merelli}, {Monet}, {Munn}, {Narayanan},
  {Nash}, {Neilsen}, {Neswold}, {Newberg}, {Nichol}, {Nicinski}, {Nonino},
  {Okada}, {Okamura}, {Ostriker}, {Owen}, {Pauls}, {Peoples}, {Peterson},
  {Petravick}, {Pier}, {Pope}, {Pordes}, {Prosapio}, {Rechenmacher}, {Quinn},
  {Richards}, {Richmond}, {Rivetta}, {Rockosi}, {Ruthmansdorfer}, {Sandford},
  {Schlegel}, {Schneider}, {Sekiguchi}, {Sergey}, {Shimasaku}, {Siegmund},
  {Smee}, {Smith}, {Snedden}, {Stone}, {Stoughton}, {Strauss}, {Stubbs},
  {SubbaRao}, {Szalay}, {Szapudi}, {Szokoly}, {Thakar}, {Tremonti}, {Tucker},
  {Uomoto}, {Vanden Berk}, {Vogeley}, {Waddell}, {Wang}, {Watanabe},
  {Weinberg}, {Yanny}, {Yasuda}, \& {SDSS Collaboration}}]{York2000}
{York}, D.~G., {Adelman}, J., {Anderson}, John~E., J., {et~al.} 2000, \aj, 120,
  1579, \dodoi{10.1086/301513}

\bibitem[{{Zhu} \& {M{\'e}nard}(2013)}]{Zhu2013}
{Zhu}, G., \& {M{\'e}nard}, B. 2013, \apj, 770, 130,
  \dodoi{10.1088/0004-637X/770/2/130}

\end{thebibliography}
\bibliographystyle{aasjournal}



\end{document}